\documentclass [12pt]{article}
\usepackage [cp1251]{inputenc}
\usepackage [english]{babel}
\usepackage {graphicx}
\usepackage {amssymb}
\usepackage {amsmath}
\usepackage {longtable}

\title{Rayleigh-Benard convection in a nonuniformly rotating electrically conductive medium in an external spiral magnetic field }
\author{$^{1}$\textbf{M.I. Kopp}, $^3$\textbf{A.V. Tur}, $^{1,2}$\textbf{V.V. Yanovsky}}

\begin{document}

\maketitle

$^{1}$ \textit{Institute for Single Crystals, NAS  Ukraine, Nauky Ave. 60, Kharkov 61001, Ukraine}

$^{2}$\textit{V.N. Karazin Kharkiv National University 4 Svobody Sq., Kharkov 61022, Ukraine}

$^{3}$\textit{Universit\'{e} de Toulouse [UPS], CNRS, Institut de Recherche en Astrophysique et Plan\'{e}tologie,
9 avenue du Colonel Roche, BP 44346, 31028 Toulouse Cedex 4, France}

\abstract{The research is devoted to the stability of convective flow in a nonuniformly rotating layer of an electrically conducting fluid  in a spiral magnetic field. The stationary and oscillatory modes of magnetic convection are considered depending on the profile of the angular rotation velocity (Rossby number $\textrm{Ro}$) and on the profile of the external azimuthal magnetic field (magnetic Rossby number $\textrm{Rb}$). The nonlinear dynamic system of Lorentz type equations is obtained by using the Galerkin method. Numerical analysis of these equations has shown the presence of chaotic behavior of convective flows. The criteria of the occurrence of chaotic movements are found. It depends on the parameters of convection: dimensionless numbers of Rayleigh $\textrm{Ra}$, Chandrasekhar $\textrm{Q}$, Taylor $\textrm{Ta}$, and external azimuthal magnetic field with the Rossby magnetic number $\textrm{Rb}=-1$ for Rayleigh $(\textrm{Ro}=-1)$ and Kepler $(\textrm{Ro}=-3/4)$ profiles of the angular rotation velocity of the medium.}


\section{Introduction}

\begin{figure}
  \centering
	\includegraphics[width=5.5 cm, height= 6 cm]{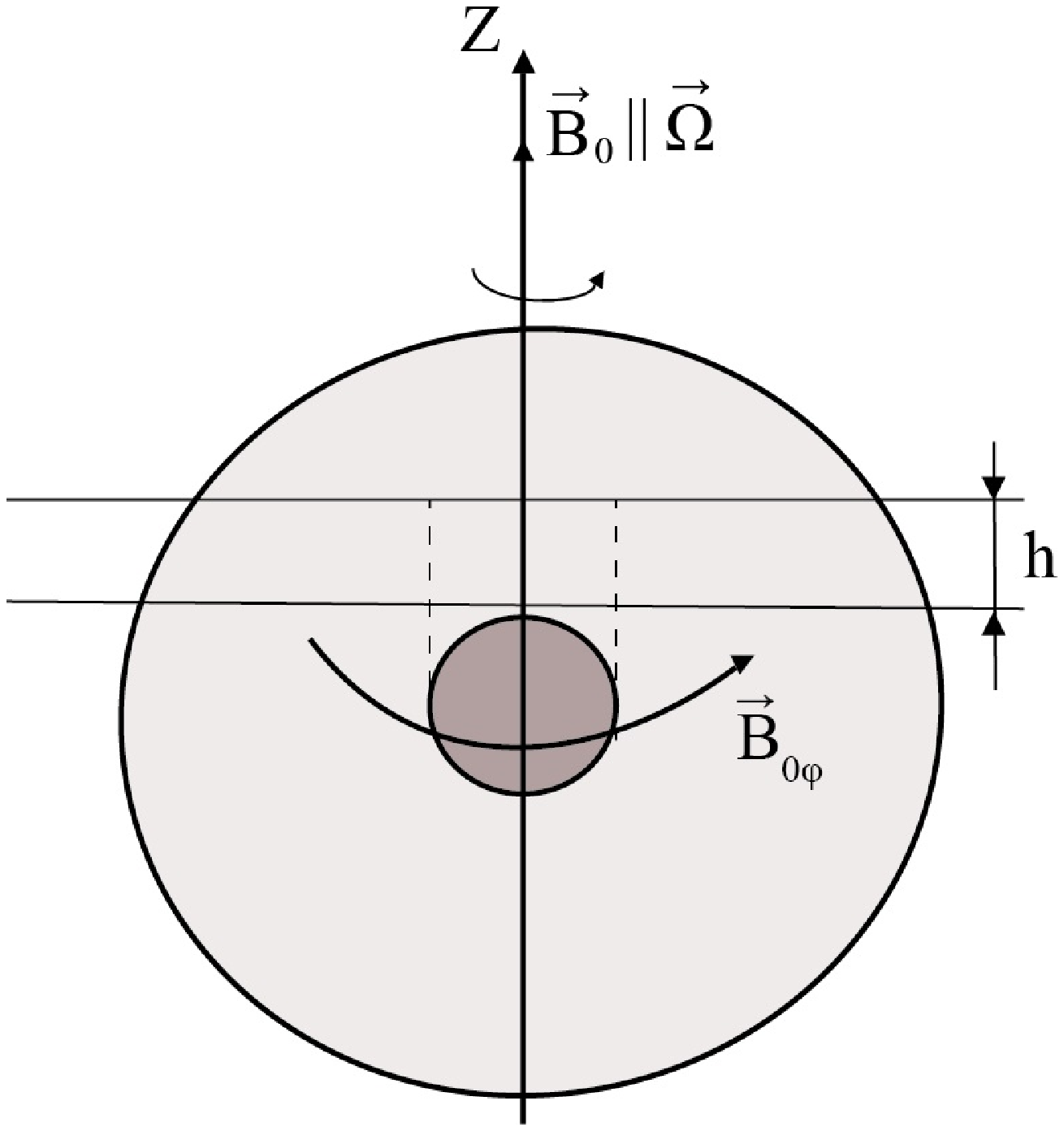}
    \includegraphics[width=6 cm, height= 6 cm]{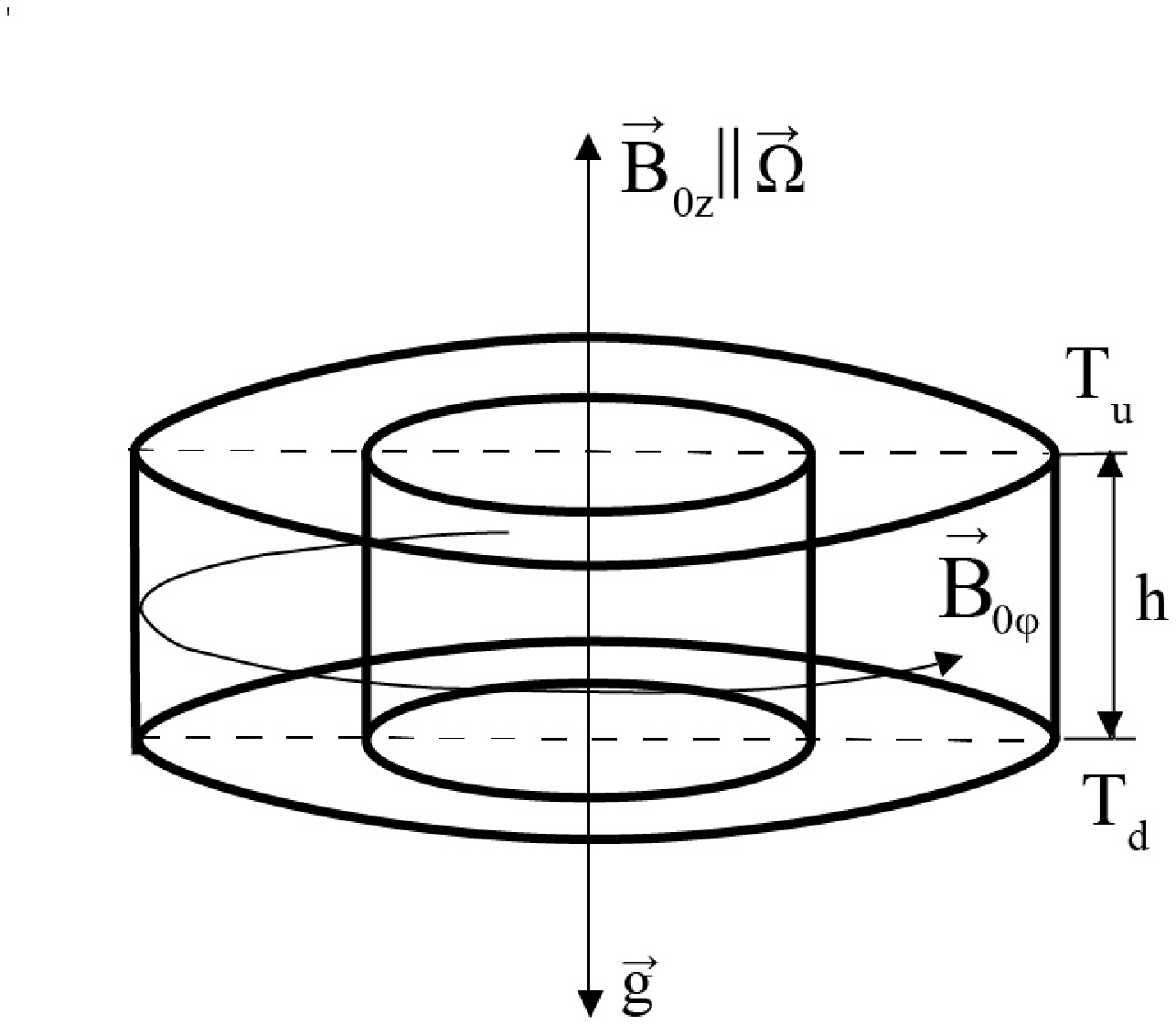}  a)-b) \\
  \caption{a) schematically shows a layer of an electrically conductive fluid of thickness $h$ of an astrophysical object, which rotates at a non-uniform velocity ${\bf \Omega}(R)$  in a spiral magnetic field $ {\bf B}_{0}=  B_{0\phi}(R){\bf e}_{\phi}+ B_{0z}{\bf e}_{z} $; b) geometric description of the Couette-Taylor flow problem in a rotating flow with a vertical temperature gradient in a spiral magnetic field.}\label{fg2}
\end{figure}

The instability of a horizontal fluid layer heated from below in the field of gravity (the Rayleigh-Benard problem) is a classic problem of fluid dynamics \cite{1s}-\cite{3s}. Problems related to the effect of rotation and magnetic field on Rayleigh-Benard convection are the most interesting, for example, because of their applications to the theory of vortex and magnetic dynamo \cite{4s}-\cite{6s}, as well as engineering and technical applications \cite{7s}. Convection, in which the axis of rotation of the medium and the uniform magnetic field coincide with the direction of the gravity vector, was well studied in \cite{1s}-\cite{2s}. The case is also interesting for astrophysical problems when the directions of the axes of rotation and the magnetic field
\begin{figure}
  \centering
	\includegraphics[width=5.5 cm, height= 5.5 cm]{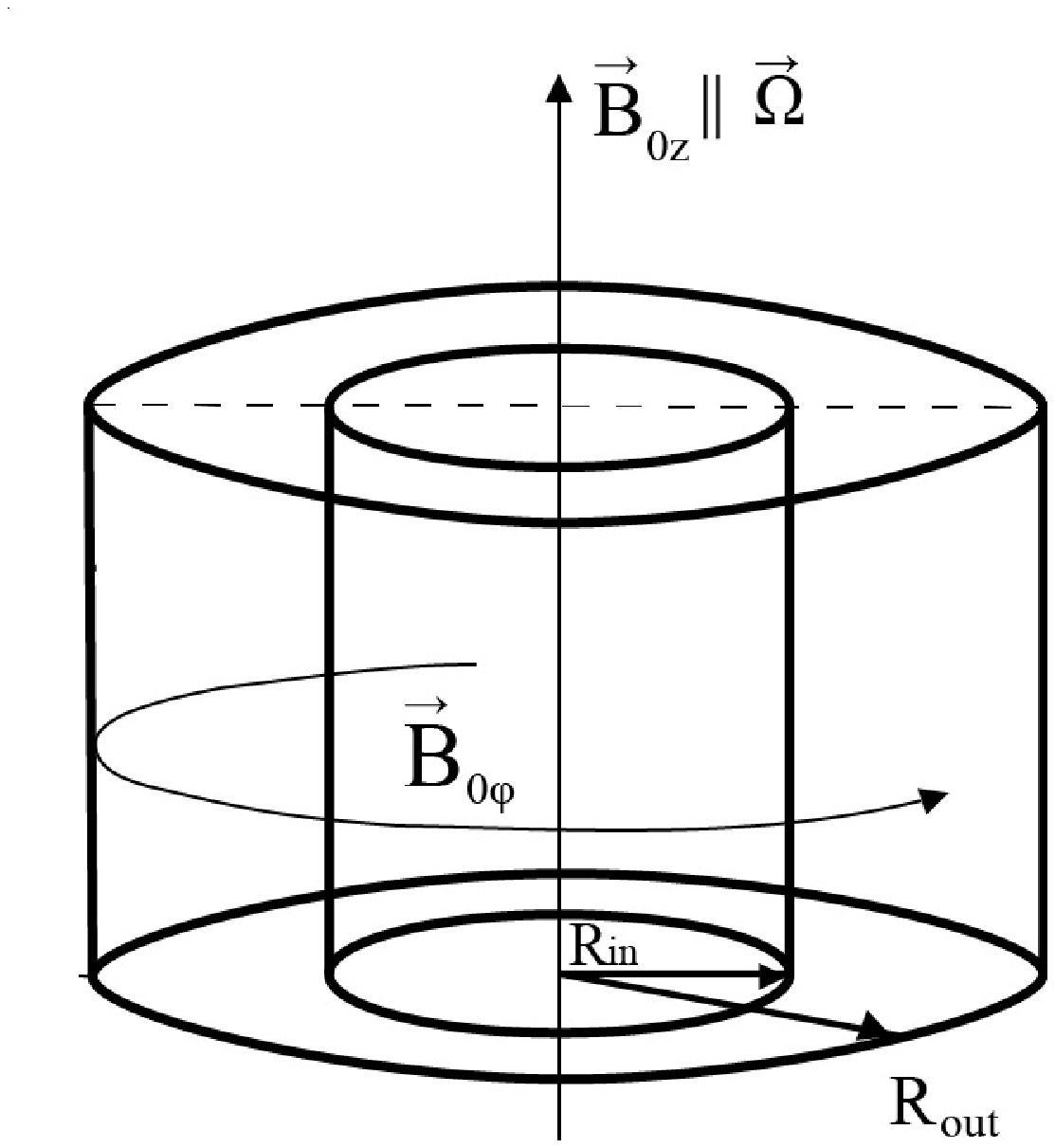}  
	\includegraphics[width=5.3 cm, height= 5.3 cm]{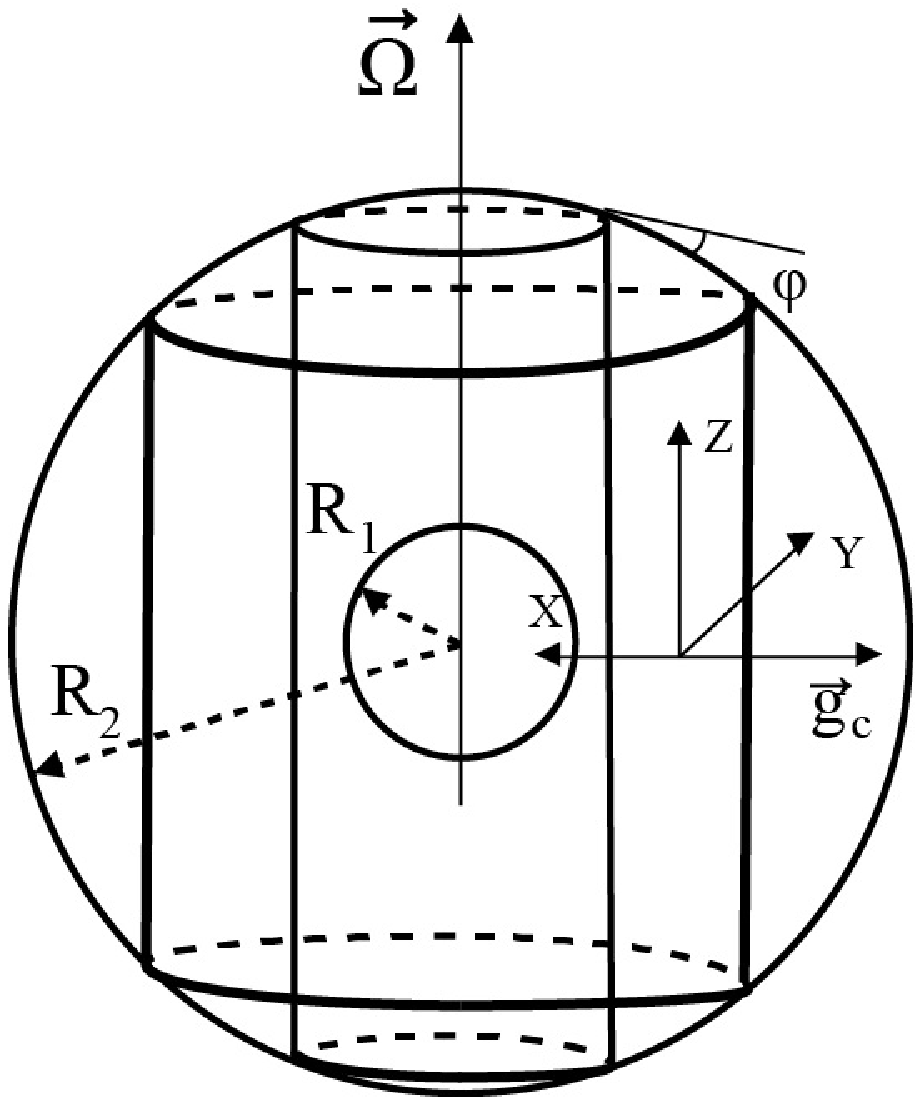}   a)-b) \\
	\caption{a) the geometry of the problem for MRI in a spiral magnetic field: two concentric cylinders with radii  $R_{\textrm{in}}=R_1$  and $R_{\textrm{out}}=R_2$  rotate at speeds $\Omega_{\textrm{in}}=\Omega_1$ and $\Omega_{\textrm{out}}=\Omega_2$. $B_{0z}$ - axial magnetic field, directed vertically upwards, $B_{0\phi}(R)$ - azimuthal or toroidal magnetic field; b) convective Busse dynamo model for a layer of an electrically conducting fluid in a rotating magnetoconvection.}\label{fg1}
\end{figure}
are perpendicular to each other, and the direction of the magnetic field is perpendicular to the direction of the gravity vector. Such problem statement corresponds to convection in fluid layers located in the equatorial region of a rotating object, where the azimuthal magnetic field plays a significant role. The linear theory of such convection was first constructed in \cite{8s}-\cite{9s}. The linear theory of rotating magnetic convection for a random deviation of the axes of rotation and the magnetic field from the vertical axis (gravity field) was developed in \cite{10s}. A weakly nonlinear theory and stability analysis of azimuthal magnetic convection with $B_{0\phi}(R)=\textrm{const}$ was performed in \cite{11s}. 
It proposes a model in which the centrifugal acceleration $g_c=\Omega^2(R_1+R_2)$ can play the role of gravitational acceleration ${\bf{g}}$ for free convection in the local Cartesian approximation. The weakly nonlinear theory of centrifugal magnetoconvection considered in \cite{11s} was applied to the problem of a hydromagnetic dynamo. In convective models of a magnetic dynamo, the generation of a magnetic field occurs due to convective flows of a conducting medium. Various models of thermal convection in rapidly rotating liquids penetrated by strong magnetic fields are discussed in \cite{12s}. An extra attention is paid to the possibility that the magnetic field can be supported by the action of a dynamo, and not by external applied electric currents. An overview of two dynamo models is given in \cite{12s}. This is the model of the Childress-Sovard flat layer \cite{13s} and the Busse annulus model \cite{14s}. The Childress-Sovard model functions in convective flat layers of fluid located in temperate and subpolar latitudes (see Fig. \ref{fg2}$\textrm{a}$ ) of a space object. For the Earth$^{,}$s dynamo, the Busse model functions in the equatorial layers, where the azimuthal magnetic field plays a significant role. Electrically conductive fluid rotates in an annular region located between the solid core and the Earth$^{}$s crust. The theory of this process was developed in \cite{14s}-\cite{15s}, where the model of rotating cylinders was used. In this theory \cite{15s}, the outer cylinder rotates with a constant angular velocity $\Omega_2$, and the inner cylinder remains stationary $\Omega_1=0$ (see Fig. \ref{fg1}$\textrm{b}$). Convective flows (Benard cells) occur in the fluid layer between the cylinders due to the temperature difference between the inner $T_{\textrm{in}}$ and the outer $T_{\textrm{out}}$ cylinder: $T_{\textrm{out}}>T_{\textrm{in}}$. The difference in heights between the inner $h_1$  and outer cylinders $h_2$ leads to a similar effect of the Coriolis force on the $\beta$-plane. This model was also actively
used to study the behavior of hydromagnetic waves in rotating magnetoconvection. As a result of the $\beta$-effect in a rotating magnetoconvection, a more general type of Rossby wave, the so-called magneto-thermal Rossby wave, arises \cite{16s}.
Recently, the weakly nonlinear stage of rotating magneto-convection has been intensively studied. In this stage a chaotic regime occurs: in rotating fluid layers \cite{17s}-\cite{18s}, in conducting mediums with a uniform magnetic field \cite{14s}-\cite{22s}, and in conducting mediums rotating with a magnetic field \cite{23s}. However, the dynamics of the magnetic field itself was not considered in these works, which corresponds to the non-inductive approximation. Such tasks have great importance for technological applications: crystal growth, chemical processes of solidification and centrifugal casting of metals, etc. The theory of convective magnetic dynamo is actively investigating issues related to the physical nature of inversions and variations of the geomagnetic field \cite{24s}-\cite{27s}. Inversions or polarity reversal of the Earth$^{,}$s magnetic field are confirmed in paleomagnetic and archaeomagnetic data (e.g., see \cite{28s}). Braginsky noted that the average geomagnetic field oscillates with a period of about $10^3$ years. Higher frequencies in the spectrum of the geomagnetic field have periods in the range of $10^2$ and shorter. The main reason for the appearance of the discrete spectrum of variations in the magnetic field is related by Braginsky to the excitation of MAC-waves caused by the action of magnetic, Archimedean and Coriolis forces \cite{29s}. The numerical calculations shown in \cite{27s} very well reflect the dipole structure of the Earth$^{,}$s magnetic field and its chaotic inversions. However, as noted in \cite{30s}, there are a number of shortcomings in the theory of convective dynamo. One of these shortcomings is associated with the introduction of overdiffusion coefficients necessary for the stability of numerical schemes. Another shortcoming is associated with ignoring the effect of the differential ( nonuniformly ) rotation of the medium, i.e. in the traditional concept of the geomagnetic dynamo, the rotation of the liquid core does not depend on the coordinates \cite{27s}.
In addition to the Earth$^{,}$s core, there are space objects that consist of dense gases or liquids (Jupiter, Saturn, Sun, etc.) that rotate non-uniformly, i.e. different parts of the object rotate around a common axis of rotation with different angular velocity. Differential rotation is also observed in galaxies, accretion disks, and extended rings of planets. It is known that such large-scale vortex structures such as typhoons, cyclones 
\begin{figure}
  \centering
	\includegraphics[width=8 cm, height= 6 cm]{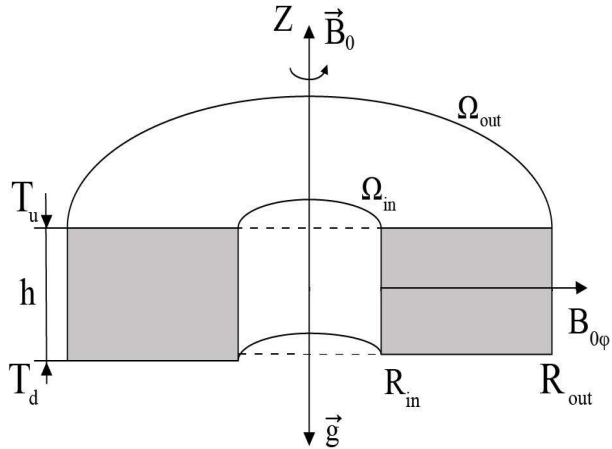} \\
\caption{The geometry of the problem for inhomogeneous rotating convection in a spiral magnetic field: ${\bf B}_{0}=  B_{0\phi}(R){\bf e}_{\phi}+ B_{0z}{\bf e}_{z} $. The electrically conductive fluid fills the layer between two rotating cylinders with angular velocities $\Omega_{\textrm{in}}$ and $\Omega_{\textrm{out}}$,  respectively. The lower surface of the layer has a temperature $T_d$, and the upper one has $T_u$: $T_{d}>T_{u}$. }\label{fg3}
\end{figure}
and anticyclones, etc. also rotate nonuniformly. The stability of the differentially rotating  plasma in an axial magnetic field  was first considered in \cite{31s}-\cite{32s}. It also shows that a weak axial magnetic field destabilizes the azimuthal differential plasma rotation, and if the condition $d \Omega^2/dR <0$ is fulfilled, a magnetorotational instability (MRI) or standard MRI (SMRI) occurs in a nondissipative plasma. Since this condition is also satisfied for the Kepler flows $\Omega \sim R^{-3/2}$, the MRI is the most possible source of turbulence in accretion disks. The discovery of the MRI was the impulse for numerous theoretical studies. The first theoretical studies considered the problem of accretion flows were performed with the approximation of a nondissipative plasma taking into account the radial thermal stratification \cite{33s}, as well as taking into account the magnetization of the heat flows \cite{34s}. The stability of differential rotating plasma in an axial magnetic field was considered with simultaneous consideration of both dissipative effects (viscosity and ohmic dissipation) and thermal radial stratification of the plasma in \cite{35s}. MRI in a spiral magnetic field, i.e. with the nontrivial topology, ${\bf{B}}_{0}\textrm{rot}{\bf{B}}_{0} \ne 0 $ was studied in \cite{36s}-\cite{37s}. In the study of MRI, the differential rotation of the medium is modeled by the Couette flow between the two cylinders rotating with different angular velocities (see Fig. \ref{fg1}$\textrm{a}$), which is convenient for implementation of laboratory experiments \cite{38s}.
The problem of the stability of an electrically conducting fluid between two rotating cylinders (Couette flow) and the Rayleigh-Benard problem in an external constant magnetic field were both considered in \cite{39s}. A study of the chaotic regime was also performed on the basis of the nonlinear dynamics equations of the six-dimensional phase space. The analysis of these equations has shown the existence of a complex chaotic structure -- a strange attractor. A convection mode in which a chaotic change in direction (inversion) and amplitude of the perturbed magnetic field, taking into account the inhomogeneous rotation of the medium, occurs was found as well.
In contrast to \cite{39s}, here we have solved the Rayleigh-Benard problem for a nonuniformly rotating layer of an electrically conducting medium (plasma) in an external spiral magnetic field (see Fig. \ref{fg2}$\textrm{b}$). The content of the work is outlined in the following sections. The equations of the evolution of small perturbations in the Boussinesq approximation in a rotating viscous incompressible electrically conducting fluid that is in a field of gravity with a constant temperature gradient are obtained in Section 2. The Rayleigh-Benard problem for a layer of an electrically conducting fluid enclosed between two rotating cylinders and heated from below is solved in Section 3. The monotone and oscillatory modes of convection are investigated in Section 4 using the dispersion equation obtained in Section 3. The critical values of Rayleigh numbers for stationary and oscillatory convective instability are also obtained in Section 4. The growth of these instabilities is analyzed for various profiles of the angular velocity of rotation $\Omega(R)$ and from the profile of the external azimuthal magnetic field $B_{0\phi}(R)$. A weakly nonlinear stage of a nonuniformly rotating convection in an external spiral magnetic field, in which a chaotic regime occurs, leading to random variations of the magnetic field, is studied for the axisymmetric perturbations in Section 5. We have obtained a dynamic system of Lorentz type equations in eight-dimensional phase space by applying the Galerkin method to a nonlinear system of equations for  nonuniformly  rotating convection in a spiral magnetic field. Analytical and numerical analysis of this system of equations has shown the chaotic behavior of the magnetic field and its inversion.
The theory developed in this work can be applied to various astrophysical and geophysical problems that consider the chaotic behavior of a magnetic field in convective layers of the Earth$^{,}$s core, the Sun, hot galactic clusters, accretion disks and other objects.

\section {Problem statement and basic evolution equations for small perturbations}

Let us consider convective flows in a nonuniformly rotating layer of an incompressible viscous electrically conducting fluid (plasma) of thickness $h$, which is enclosed between two rotating cylinders with an inner radius $R_{\textrm{in}}$ and outer radius $R_{\textrm{out}}$, and $h \ll (R_{\textrm{out}}-R_{\textrm{in}})$. The temperature $T_{d}$ of the lower plane of the layer is maintained to be higher than the temperature of the upper plane $T_{u}$: $T_{d}>T_{u}$ -- heated from below (see Fig. \ref{fg3}). We assume that the fluid is in a constant gravitational field ${\bf{g}}$, directed along the $OZ$ axis vertically downwards: ${\bf{g}}=(0,0,-g)$. The electrically conductive fluid rotates at the angular velocity ${\bf{\Omega} }$ directed vertically upward along the axis $OZ$. The rotation of the fluid creates a stationary  flow in the azimuthal direction: ${\bf v}_0={\bf e}_\phi\Omega (R)R$, where $\Omega (R)$ is the angular velocity of rotation with an arbitrary dependence on the coordinate $R$. In addition, we assume that the plasma is in a spiral magnetic field ${\bf B}_{0}$, which we represent as the sum of a non-uniform azimuthal $ B_{0\phi}(R)$ and a uniform axial $ B_{0z}$ fields:
\[{\bf B}_{0}=  B_{0\phi}(R){\bf e}_{\phi}+ B_{0z}{\bf e}_{z},\quad B_{0z}=\textrm{const}, \quad {\bf B}_{0} \textrm{rot} {\bf B}_{0} \neq 0  .\]
Such geometry of the problem generalizes the classical Rayleigh-Benard problem for free convection. Obviously, it is convenient to use a cylindrical coordinate system $(R,\phi ,z)$, chosen according to the possibility of practical application of the theory developed in this paper, to study this type of flow. Then the equations of motion of a viscous incompressible electrically conducting fluid in the Boussinesq approximation for a cylindrical coordinate system take the following form \cite{2s}:
\begin{equation} \label{eq1} \frac{\partial v_{R} }{\partial t} +({\bf v}\nabla)v_R-\frac{v_\phi^2}{R} -\frac{1}{4\pi \rho _{0} } \left(({\bf B}\nabla)B_R-\frac{B_\phi^2}{R}\right) =-\frac{1}{\rho _{0} } \frac{\partial }{\partial R}\left(P+\frac{B^2}{8\pi}\right) +$$
$$+\nu \left(\nabla^2 v_{R} -\frac{2}{R^2}\frac{\partial v_\phi}{\partial \phi} - \frac{v_R}{R^2} \right)\end{equation}
\begin{equation} \label{eq2} \frac{\partial v_{\phi } }{\partial t} + ({\bf v}\nabla)v_\phi+\frac{v_\phi v_R}{R}-\frac{1}{4\pi\rho_{0}}\left( ({\bf B}\nabla)B_\phi+ \frac{B_\phi B_R}{R}\right)=-\frac{1}{\rho_{0} R} \frac{\partial}{\partial \phi }\left(P+\frac{B^2}{8\pi}\right) +$$
$$+\nu  \left(\nabla^2 v_{\phi} +\frac{2}{R^2}\frac{\partial v_R}{\partial \phi} - \frac{v_\phi}{R^2} \right)  \end{equation} 
\begin{equation} \label{eq3} \frac{\partial v_{z}}{\partial t}+({\bf v}\nabla)v_z -\frac{1}{4\pi\rho_{0}}({\bf B}\nabla)B_z=-\frac{1}{\rho _{0} } \frac{\partial }{\partial z}\left(P+\frac{B^2}{8\pi}\right) + g\beta T +\nu \nabla^2 u_{z}  \end{equation}
\begin{equation} \label{eq4} \frac{\partial B_{R} }{\partial t}+({\bf v}\nabla)B_R-({\bf B}\nabla)v_R=\eta \left(\nabla^2 B_{R} -\frac{2}{R^2}\frac{\partial B_\phi}{\partial \phi} - \frac{B_R}{R^2} \right)  \end{equation} 
\begin{equation} \label{eq5} \frac{\partial B_{\phi } }{\partial t}+({\bf v}\nabla)B_\phi-({\bf B}\nabla)v_\phi +\frac{1}{R}\left(v_\phi B_R-v_RB_\phi\right)-=$$
$$=\eta\left(\nabla^2 B_{\phi} +\frac{2}{R^2}\frac{\partial B_R}{\partial \phi} - \frac{B_\phi}{R^2} \right)   \end{equation} 
\begin{equation} \label{eq6} \frac{\partial B_{z} }{\partial t} +({\bf v}\nabla)B_z-({\bf B}\nabla)v_z  =\eta \nabla^2 B_{z}  \end{equation} 
\begin{equation} \label{eq7} \frac{\partial T }{\partial t} + ({\bf v}\nabla)T =\chi \nabla^2 T  \end{equation} 
\begin{equation} \label{eq8} \frac{\partial v_{R} }{\partial R} +\frac{v_{R} }{R} +\frac{1}{R} \frac{\partial v_{\phi } }{\partial \phi } +\frac{\partial v_{z} }{\partial z} =0, $$
$$ \frac{\partial B_{R} }{\partial R} +\frac{B_{R} }{R} +\frac{1}{R} \frac{\partial B_{\phi } }{\partial \phi } +\frac{\partial B_{z} }{\partial z} =0,  \end{equation} 
where the scalar product $({\bf A}\nabla)$ and Laplacian $\Delta\equiv\nabla^2$  are respectively equal to
\[ ({\bf A}\nabla)=A_R\,\frac{\partial}{\partial R}+\frac{A_\phi}{R}\frac{\partial}{\partial\phi}+A_z\,\frac{\partial}{\partial z}, \; \Delta=\frac{\partial^2}{\partial R^2}+\frac{1}{R}\frac{\partial}{\partial R}+\frac{1}{R^2}\frac{\partial^2}{\partial\phi^2}+\frac{\partial^2}{\partial z^2} ,\]
where $\beta $ is the coefficient of thermal expansion, $\rho _{0} =\textrm{const}$  is the density of the medium, $\nu $ is the kinematic viscosity, $\eta=c^2/4\pi\sigma $ is the magnetic viscosity, $\sigma $ is the electrical conductivity coefficient, and $\chi $  is the thermal conductivity coefficient of the medium.
Let us represent all the quantities in the equations (\ref{eq1})-(\ref{eq8}) as the sum of the stationary and perturbed parts:
\begin{equation} \label{eq9}
 \begin{pmatrix}  v_R \\ v_\phi \\ v_z \end{pmatrix}=\begin{pmatrix}  0 \\ \Omega(R)R \\ 0 \end{pmatrix}+ \begin{pmatrix}  u_R \\ u_\phi \\ u_z \end{pmatrix}, \begin{pmatrix}  B_R \\ B_\phi \\ B_z \end{pmatrix}=\begin{pmatrix}  0 \\ B_{0\phi} \\ B_{0z} \end{pmatrix}+ \begin{pmatrix}  b_R \\ b_\phi \\ b_z \end{pmatrix}, P=p_0+p, T=T_0+\theta . \end{equation}
The stationary flow equilibrium is ensured by the balance of forces in the radial direction:
\begin{equation} \label{eq10}
 \Omega ^{2} R=\frac{1}{\rho _{0} } \frac{dp_{0} }{dR}+\frac{B_{0\phi}}{4\pi\rho_0 R} \frac{d}{dR}(R B_{0\phi}) \end{equation} 
The hydrostatics equation satisfies the steady state in the vertical direction:
\begin{equation} \label{eq11}
 \frac{1}{\rho _{0} } \frac{dp_{0} }{dz} =-g\beta T_{0} \end{equation} 
The stationary temperature profile  $T_0=T_0(z)$ is found by solving the following equation
\begin{equation} \label{eq11a} 
 \nabla^2 T_0=0,\quad  \textrm{at}\quad T(0)=T_d,\; T(h)=T_u .  \end{equation}
The system of equations (\ref{eq1})-(\ref{eq8}) allows stationary solutions of the general form:
$$P=p_{0} (R,z),\;\rho=\rho_0=\textrm{const},\; B_{R}(R)=0,\; B_{\phi }=B_{0\phi }(R), $$
$$ B_{z}(R)=B_{0} =\textrm{const},\; v_{R} (R)=0, \; v_{\phi }(R)=R\Omega(R),$$
$$ v_{z} (R)=0,\; T_{0}(z) =T_d-\frac{T_d-T_u}{h}\cdot z. $$
Further considerations include the issue of the stability of small perturbations of physical quantities $({\bf{u}}=(u_R,u_\phi,u_z), {\bf{b}}=(b_R,b_\phi,b_z), p, \theta)$ on the background of the stationary state. The evolution equations for small perturbations in the linear approximation are obtained by substituting  (\ref{eq9}) into (\ref{eq1})-(\ref{eq8}):
\begin{equation} \label{eq12} \frac{\partial u_{R} }{\partial t} +\Omega \frac{\partial u_{R} }{\partial \phi } -2\Omega u_{\phi } -\frac{1}{4\pi \rho _{0} }\left(\frac{B_{0\phi}}{R}\frac{\partial b_R}{\partial \phi}-\frac{2B_{0\phi}b_\phi}{R}+ B_{0} \frac{\partial b_{R} }{\partial z}\right) =$$
$$=-\frac{1}{\rho _{0} } \frac{\partial \widetilde p}{\partial R} +\nu \left(\nabla^2 u_{R} -\frac{2}{R^2}\frac{\partial u_\phi}{\partial \phi} - \frac{u_R}{R^2} \right)\end{equation} 
\begin{equation} \label{eq13} \frac{\partial u_{\phi } }{\partial t} +\Omega \frac{\partial u_{\phi } }{\partial \phi } +2\Omega(1+\textrm{Ro})u_{R} -\frac{1}{4\pi \rho _{0} }\left(\frac{B_{0\phi}}{R}\frac{\partial b_R}{\partial \phi}+\frac{2B_{0\phi}}{R}\left(1+\textrm{Rb}\right)+B_{0} \frac{\partial b_{\phi } }{\partial z}\right) =$$
$$=-\frac{1}{\rho_{0} R} \frac{\partial \widetilde p}{\partial \phi } +\nu  \left(\nabla^2 u_{\phi} +\frac{2}{R^2}\frac{\partial u_R}{\partial \phi} - \frac{u_\phi}{R^2} \right)  \end{equation} 
\begin{equation} \label{eq14} \frac{\partial u_{z} }{\partial t} +\Omega \frac{\partial u_{z} }{\partial \phi } -\frac{1}{4\pi \rho _{0} }\left(\frac{B_{0\phi}}{R}\frac{\partial b_z}{\partial \phi}+ B_{0} \frac{\partial b_{z} }{\partial z}\right) =-\frac{1}{\rho _{0} } \frac{\partial \widetilde p}{\partial z} + g\beta \theta +\nu \nabla^2 u_{z}  \end{equation} 
\begin{equation} \label{eq15} \frac{\partial b_{R} }{\partial t} +\Omega \frac{\partial b_{R} }{\partial \phi }-\frac{B_{0\phi}}{R}\frac{\partial u_R}{\partial \phi} -B_{0} \frac{\partial u_{R} }{\partial z} =\eta \left(\nabla^2 b_{R} -\frac{2}{R^2}\frac{\partial b_\phi}{\partial \phi} - \frac{b_R}{R^2} \right)  \end{equation} 
\begin{equation} \label{eq16} \frac{\partial b_{\phi } }{\partial t} +\Omega \frac{\partial b_{\phi } }{\partial \phi }-\frac{B_{0\phi}}{R}\frac{\partial u_\phi}{\partial \phi}-B_{0} \frac{\partial u_{\phi } }{\partial z}-2\Omega \textrm{Ro} b_{R} + \frac{2B_{0\phi}}{R}\textrm{Rb} u_R=$$
$$=\eta\left(\nabla^2 b_{\phi} +\frac{2}{R^2}\frac{\partial b_R}{\partial \phi} - \frac{b_\phi}{R^2} \right)   \end{equation} 
\begin{equation} \label{eq17} \frac{\partial b_{z} }{\partial t} +\Omega \frac{\partial b_{z} }{\partial \phi }-\frac{B_{0\phi}}{R}\frac{\partial u_z}{\partial \phi}-B_{0} \frac{\partial u_{z} }{\partial z} =\eta \nabla^2 b_{z}  \end{equation} 
\begin{equation} \label{eq18} \frac{\partial \theta }{\partial t} +\Omega \frac{\partial \theta }{\partial \phi } -u_{z}\left(\frac{T_d-T_u}{h}\right)=\chi \nabla^2 \theta  \end{equation} 
\begin{equation} \label{eq19} \frac{\partial u_{R} }{\partial R} +\frac{u_{R} }{R} +\frac{1}{R} \frac{\partial u_{\phi } }{\partial \phi } +\frac{\partial u_{z} }{\partial z} =0, $$
$$ \frac{\partial b_{R} }{\partial R} +\frac{b_{R} }{R} +\frac{1}{R} \frac{\partial b_{\phi } }{\partial \phi } +\frac{\partial b_{z} }{\partial z} =0, 
\end{equation} 
where $ \widetilde p=p+\frac{1}{{4\pi }}({\bf B}_0 {\bf b})$ is the total perturbed pressure, $\textrm{Ro}=\frac{R}{2\Omega} \frac{\partial\Omega}{\partial R}$  is the Rossby hydrodynamic number \cite{37s} that characterizes the inhomogeneity of the medium rotation, $\textrm{Rb}=\frac{R}{2B_{0\phi}R^{-1}}\frac{\partial}{\partial R}(B_{0\phi}R^{-1})$  is the Rossby magnetic number \cite{37s} that characterizes the inhomogeneity of the magnetic field. Note that the Rossby parameter is zero $\textrm{Ro}=0$ for solid body rotation, $\textrm{Ro}=-3/4$ in the case of Keplerian rotation, and $\textrm{Ro}=-1$ for the Rayleigh profile of the angular velocity $ \Omega(R)\sim R^{-2}$ respectively. The azimuthal component of the magnetic field  ${\bf{B}}_{0}$  can be created by an external axial current $I$: $B_{0\phi}(R)= 2I/R$. The magnetic Rossby number is $\textrm{Rb}=-1$ for the dependency $B_{0\phi}\sim R^{-1}$. The case in which $\textrm{Rb}=0$  is the result of a linear dependency of the magnetic field on the radial direction $B_{0\phi}(R)\sim R$, since it will be created by a uniform axial current in the fluid.

\section  {Rayleigh-Benard problem for a thin layer of a nonuniformy rotating magnetized plasma}

The system of equations (\ref{eq12})-(\ref{eq19}) will be used to study the issue of the stability of small perturbations. Since the characteristic scale of medium inhomogeneity in the horizontal plane is larger than in the vertical direction $L_{R}\gg L_{h}$, we can apply the local WKB method for perturbations that depend on the horizontal coordinates $(R, \phi)$. For this purpose, we have expanded all the quantities into a Taylor series in an area of fixed points $(R_0, \phi_0)$ and left the zero order members at local coordinates $\widetilde{R}=R-R_0$, $\widetilde{\phi}=\phi-\phi_0$. As a result, a system of differential equations with constant coefficients is obtained by taking into account the following relations
$$\Omega_0=\Omega(R_0),\quad \nabla^2 \rightarrow \widehat{D}^2+\frac{\partial^2}{\partial\widetilde{R}^2}+\frac{1}{R_0}\frac{\partial}{\partial \widetilde{R}}+\frac{1}{R_0^2}\frac{\partial}{\partial\widetilde{\phi}^2}, \quad \widehat D \equiv \frac{\partial}{\partial z} ,  $$
$$ \left(\nabla^2 \begin{pmatrix} \bf u \\ \bf b \end{pmatrix}\right)_{R}= \nabla^2 \begin{pmatrix}  u_R \\ b_R \end{pmatrix}-\frac{2}{R_0^2}\frac{\partial }{\partial {\phi}}\begin{pmatrix}  u_\phi \\ b_\phi \end{pmatrix}-\frac{1}{R_0^2}\begin{pmatrix}  u_R \\ b_R \end{pmatrix}, $$
$$ \left(\nabla^2 \begin{pmatrix} \bf u \\ \bf b \end{pmatrix}\right)_{\phi}= \nabla^2 \begin{pmatrix}  u_\phi \\ b_\phi \end{pmatrix}+\frac{2}{R_0^2}\frac{\partial }{\partial {\phi}}\begin{pmatrix}  u_R \\ b_R \end{pmatrix}-\frac{1}{R_0^2}\begin{pmatrix}  u_\phi \\ b_\phi \end{pmatrix} . $$
All perturbations in the system of equations (\ref{eq12})-(\ref{eq19}) can be represented in the form of plane waves
\begin{equation} \label{eq20}
\begin{pmatrix} \bf u \\ \bf b \\ \theta \\ \widetilde p \end{pmatrix}= \begin{pmatrix} {\bf U}(z) \\ {\bf H}(z) \\ \Theta(z) \\ \widetilde P(z) \end{pmatrix} \exp (\gamma t + im\widetilde{\phi } + ik \widetilde{R})
\end{equation} 
The following is found after substituting (\ref{eq20}) into the system of equations (\ref{eq12})-(\ref{eq19}), in the shortwave approximation $ k \gg \frac{1}{R_0}$, neglecting the terms  $\frac{ik}{R_0},  -\frac{1}{R_0^2} $:
\begin{equation} \label{eq21}
\widehat L_{\nu} U_R  + \frac{{2\Omega }}{\nu }U_\phi-\frac{2im}{\nu R_0^2}U_\phi+ \frac{{imB_{0\phi } }}{{4\pi \rho _0 \nu R_0}}H_R  - \frac{{B_{0\phi } }}{{2\pi \rho _0 \nu R_0 }}H_\phi   + \frac{{B_{0z} }}{{4\pi \rho _0 \nu }}\widehat D H_R  - \frac{{ik }}{{\nu \rho _0 }}\widetilde P = 0 
\end{equation}
 \begin{equation} \label{eq22}
\widehat L_{\nu} U_\phi  - \frac{{2\Omega }}{\nu }(1+\textrm{Ro})U_R+\frac{2im}{\nu R_0^2}U_R+ \frac{{imB_{0\phi } }}{{4\pi \rho _0 \nu R_0 }}H_\phi  + \frac{{B_{0\phi } }}{{2\pi \rho _0 \nu R_0}}(1+\textrm{Rb})H_R + \frac{{B_{0z} }}{{4\pi \rho _0 \nu }}\widehat D H_\phi  - \frac{{im }}{{\nu  \rho _0 R_0 }}\widetilde P = 0 
\end{equation}
\begin{equation} \label{eq23}
\widehat L_{\nu} U_z + \frac{{imB_{0\phi } }}{{4\pi \rho _0 \nu R_0}}H_z + \frac{{B_{0z} }}{{4\pi \rho _0 \nu }}\widehat D H_z +\frac{g\beta}{\nu} \Theta- \frac{{\widehat D}}{{\nu \rho _0 }}\widetilde P = 0 
\end{equation}
\begin{equation} \label{eq24}
\widehat L_{\eta} H_R   + \frac{{imB_{0\phi } }}{{ \eta R_0 }}U_R+ \frac{{B_{0z} }}{{\eta }}\widehat D U_R = 0 
\end{equation}
\begin{equation} \label{eq25}
\widehat L_{\eta} H_\phi   + \frac{{imB_{0\phi } }}{{ \eta R_0 }}U_\phi+ \frac{{B_{0z} }}{{\eta }}\widehat D U_\phi-\frac{2B_{0\phi}}{\eta R_0} \textrm{Rb} U_R+ \frac{2\Omega}{\eta} \textrm{Ro} H_R= 0 
\end{equation}
\begin{equation} \label{eq26}
\widehat L_{\eta} H_z   + \frac{{imB_{0\phi } }}{{ \eta R_0 }}U_z+ \frac{{B_{0z} }}{{\eta }}\widehat D U_z = 0 
\end{equation}
\begin{equation} \label{eq27}
\widehat L_{\chi}\Theta+\frac{A}{\chi} U_z = 0 
\end{equation}
The following notations for operators are introduced
\[\widehat L_{\nu}=\widehat D^2-\left( {\frac{{\gamma  + m\Omega }}{\nu } + k^2+\frac{m^2}{R_0^2} } \right), \widehat L_{\eta}=\widehat D^2-\left( {\frac{{\gamma  + m\Omega }}{\eta } + k^2+\frac{m^2}{R_0^2} } \right),\]
\[ \widehat L_{\chi}=\widehat D^2-\left( {\frac{{\gamma  + m\Omega }}{\chi } + k^2+\frac{m^2}{R_0^2} } \right).\]
For the subsequent analysis of the system of equations (\ref{eq21})-(\ref{eq27}), it is convenient to bring it to a dimensionless form by introducing dimensionless quantities, which we have noted with an asterisk:
\[\left( {R_0}^ * , z^ * \right) = h^{-1} (R_0,z), \left( {U_R ^ *  ,U_\phi ^ *  ,U_z ^ *  } \right) = \chi h^{ - 1} (U_R ,U_\phi  ,U_z ), \left( {H_R ^ *  ,H_\phi ^ *  ,H_z ^ *  } \right) = B_0^{ - 1} (H_R ,H_\phi  ,H_z ), \]
\[\phi^* =\phi,\quad \Theta ^ *   = \Theta (Ah)^{ - 1},\quad  P^ *   = P\left( {\frac{{h^2 }}{{\rho _0 \nu \chi }}} \right) ,\quad t^ *   = t\left( {\frac{\nu }{{h^2 }}} \right), \quad \frac{\partial }{{\partial t^ *  }} = \frac{{h^2 }}{\nu }\frac{\partial }{{\partial t}}.\]
Omitting the asterisk icon, we have got the following system of dimensionless equations
 \begin{equation} \label{eq28}
\widehat L_\nu  U_R  + \sqrt {\textrm{Ta}} U_\phi-\frac{2im}{R_0^2}U_\phi+\Pr \textrm{Pm}^{-1}\textrm{Ha}^2 \xi (imH_R-2H_\phi) + \Pr \textrm{Pm}^{-1} \textrm{Ha}^2 \widehat DH_R  - ik \widetilde P = 0 \end{equation}
\begin{equation} \label{eq29}
\widehat L_\nu  U_\phi   - \sqrt {\textrm{Ta}} (1 +\textrm{Ro})U_R+\frac{2im}{R_0^2}U_R  + \Pr \textrm {Pm}^{-1}\textrm{Ha}^2 \xi (imH_\phi   + 2(1 +\textrm{Rb})H_R ) +$$
$$+\Pr \textrm{Pm}^{-1}\textrm{Ha}^2 \widehat DH_\phi-\frac{{im}}{R_0}\widetilde P = 0 \end{equation}
\begin{equation} \label{eq30}
\widehat L_\nu  U_z  + \Pr \textrm{Pm}^{-1}\textrm{Ha}^2 \xi imH_z  + \Pr\textrm{Pm}^{-1}\textrm{Ha}^2 \widehat DH_z  +\textrm{Ra}\Theta  - \widehat D\widetilde P = 0 \end{equation}
\begin{equation} \label{eq31}
\widehat L_\eta  H_R  + {\Pr}^{-1}\textrm{ Pm}\xi imU_R  + {\Pr}^{-1}\textrm{Pm} \widehat DU_R  = 0 \end{equation}
\begin{equation} \label{eq32}
\widehat L_\eta  H_\phi   + {\Pr}^{-1}\textrm{Pm}\xi imU_\phi+ {\Pr}^{-1}\textrm{Pm}\widehat DU_\phi-2{\Pr}^{-1}\textrm{ Pm}\xi \textrm{Rb}U_R  +\textrm{PmRo}\sqrt{\textrm{Ta}}H_R  = 0 \end{equation}
\begin{equation} \label{eq33}
\widehat L_\eta  H_z  + {\Pr}^{-1}\textrm{Pm}\xi imU_z  + {\Pr}^{-1}\textrm{Pm}\widehat DU_z  = 0 \end{equation}
\begin{equation} \label{eq34}
\widehat L_\chi \Theta+ U_z  = 0, \end{equation}
\[\widehat L_\nu   = \widehat D^2  - \gamma  - im \frac{{\sqrt {\textrm{Ta}}}}{2} - k^2-\frac{m^2}{R_0^2}, \]
\[ \widehat L_\eta   = \widehat D^2  -\textrm{Pm}\left( {\gamma- im\frac{{\sqrt {\textrm{Ta}} }}{2}} \right)-k^2-\frac{m^2}{R_0^2} ,\]
\[  \widehat L_\chi   = \widehat D^2 -\Pr \left( {\gamma-im\frac{{\sqrt {\textrm{Ta}} }}{2}} \right) - k^2-\frac{m^2}{R_0^2}, \]
where dimensionless parameters: $\textrm{Ta}=\frac{4{\Omega}^2 h^4}{\nu^2}$  is the Taylor number, $\textrm{Pr}=\nu/\chi$  is the Prandtl number, $\textrm{Pm}=\nu/\eta$  is the Prandtl magnetic number, $\textrm{Ha}=\frac{B_0 h}{\sqrt{4\pi\rho_0\nu\eta}}$  is the Hartman number, $\xi= \frac{B_{0\phi}}{R_0 B_0}$ is the ratio between the azimuth magnetic field and axial field, $\textrm{Ra}=\frac{g\beta A h^4}{\nu\chi}$  is the Rayleigh number. The Chandrasekhar number $\textrm{Q}=\textrm{Ha}^2$ will be used in future instead of the Hartmann number $\textrm{Ha}$.

Equations (\ref{eq28})-(\ref{eq34}) are supplemented by the solenoid equations of the fields ${\bf u}$ and ${\bf b}$:
\begin{equation} \label{eq35}
\widehat D U_z+ikU_R+\frac{im}{R_0}U_\phi=0,\quad \widehat D H_z+ikH_R+\frac{im}{R_0}H_\phi=0. \end{equation}
Let us consider the evolution of axisymmetric perturbations, i.e. not dependent on the azimuth angle  $\phi$ $(m=0)$. Using condition (\ref{eq35}), we exclude the pressure $\widetilde P$ from equations (\ref{eq28}) and (\ref{eq30}):
\begin{equation} \label{eq36} 
\widetilde P=\frac{1}{\widehat D^2-k^2}\left[\sqrt {\textrm{Ta}}ikU_\phi- \textrm{Ra}\frac{\widehat D U_z}{\widehat L_\chi}-2 \textrm{Q}\xi ik\left(-\frac{\widehat D U_\phi}{\widehat L_\eta}+2\xi\textrm{Rb}\frac{U_R}{\widehat L_\eta}+\textrm{PmRo}\sqrt{\textrm{Ta}}\frac{\widehat D}{\widehat L_\eta^2}U_R\right)\right]
\end{equation}
As a result of simple but bulky transformations, we have obtained one differential equation for $U_z$ by substituting the expression (\ref{eq36}) into the system of linear equations (\ref{eq28})-(\ref{eq34}):
\begin{equation} \label{eq37}
\left[ {\widehat a_{33} \left( {\widehat a_{11} \widehat a_{22}  - \widehat a_{21} \widehat a_{12} } \right) + \widehat a_{13} \left( {\widehat a_{21} \widehat a_{32}  - \widehat a_{31} \widehat a_{22} } \right)} \right]U_z  = 0, \end{equation}
where
\[\widehat a_{11}  = \widehat L_\nu   - \textrm{Q} \frac{{\widehat D^2 }}{{\widehat L_\eta}}- 2\textrm {Q}\xi\left(2\xi\textrm{Rb} +\sqrt{\textrm{Ta} }\textrm{RoPm}\frac{\widehat D}{\widehat L_\eta}\right) \cdot \frac{\widehat D^2}{\widehat L_\eta(\widehat D^2- k^2)},\; \widehat a_{12}  =\frac{\widehat D^2 }{\widehat D^2-k^2}\cdot\left( \sqrt {\textrm{Ta}}+2\textrm{Q}\xi\frac{\widehat D}{\widehat L_\eta} \right),\]
\[  \widehat a_{13}  = \frac{ik\textrm{Ra}\widehat D}{(\widehat D^2-k^2 )\widehat L_\chi},\; \widehat a_{21}=-\sqrt{\textrm{Ta}}(1 +\textrm{Ro})+\textrm{QRoPm}\sqrt {\textrm{Ta}}\frac{\widehat D^2}{\widehat L_\eta^2}-2\textrm{Q}\xi\frac{\widehat D}{\widehat L_\eta}, \]
\[\widehat a_{22}=\widehat L_\nu-\textrm{Q}\frac{\widehat D^2}{\widehat L_\eta},\; \widehat a_{31}=\frac{2\textrm {Q}\xi ik}{\widehat D^2-k^2}\cdot\left(2\xi\textrm{Rb}\frac{\widehat D}{\widehat L_\eta}+\sqrt{\textrm{Ta}}\textrm{RoPm}\frac{\widehat D^2}{\widehat L_\eta^2}\right), \]
\[\widehat a_{32}=-\frac{ik \widehat D}{\widehat D^2-k^2}\cdot\left(\sqrt{\textrm{Ta}}+2\textrm {Q}\xi\frac{\widehat D}{\widehat L_\eta}\right),\;
\widehat a_{33}  = \widehat L_\nu-\textrm {Q}\frac{\widehat D^2}{\widehat L_\eta}+\frac{k^2\textrm{Ra}}{\widehat L_\chi(\widehat D^2-k^2)}. \]
Equation (\ref{eq37}) is supplemented by boundary conditions only in the $z$-direction
\begin{equation} \label{eq38}
U_z=\frac{d^2 U_z}{dz^2}=0, \quad \textrm{at} \quad z=0,\quad z=1. \end{equation}
Equation (\ref{eq37}) describes convective phenomena in a thin layer of a nonuniformly rotating electrically conducting fluid in an external spiral magnetic field. Let us continue with a more detailed analysis of this equation.

\section {Monotonic and oscillatory modes of convection}

Let us choose the function $U_z$, which satisfies the free boundary conditions (\ref{eq38}), in the following form
\begin{equation} \label{eq39} U_z=U_{0z} \sin n\pi z \quad (n=1,2,3 \ldots), \end{equation}
where $U_{0z}=\textrm{const}$  is the disturbance amplitude, $z$ are the velocity components. Substituting (\ref{eq39}) into (\ref{eq37}), and integrating over the layer thickness $z=(0,1)$, we obtain the dispersion equation for the single-mode approximation ($n=1$):
\begin{equation} \label{eq40} 
 \textrm{Ra} = \frac{\Gamma_\chi (a^2\Gamma_A^4+\pi^2\textrm{Ta}(1+\textrm{Ro})\Gamma_\eta^2+\pi^4\textrm{QTaRoPm}-4\pi^4\textrm{Q}^2\xi^2-4\pi^2\textrm{Q}\xi^2\Gamma_A^2\textrm{Rb})}{k^2\Gamma_\eta\Gamma_A^2}  , \end{equation}
where
\[\Gamma_A^2  = (\gamma  + a^2 )(\gamma\textrm{Pm} + a^2 ) + \pi^2\textrm{Q},\quad \Gamma_\chi   = \gamma \Pr+a^2, \quad  
\Gamma_\eta   = \gamma\textrm{Pm}+a^2, \quad a^2=\pi^2+k^2.  \]
The dispersion equation (\ref{eq40}) was obtained in \cite{39s} for the case when the azimuthal magnetic field is missing $\xi=0$. Excluding thermal processes ($\textrm{Ra}=0$) and $\xi=0$, equation (\ref{eq40}) coincides with the dispersion equation for the standard MRI (SMRI) with considering dissipative processes \cite{36s}-\cite{37s}. The threshold value of the hydrodynamic Rossby number $\textrm{Ro}$ is determined using the condition  $\gamma=0$ and has the form:
\[\textrm{Ro}_{\textrm{cr}}  =-\frac{a^2(a^4+\pi^2 \textrm{Q})^2+\pi^2a^4 \textrm{Ta}}{\pi^2\textrm{Ta}(a^4+\pi^2 \textrm{Q} \textrm{Pm})} \]
When transited to dimensional variables
\[\frac{\pi^2 \textrm{Q}}{a^4} \rightarrow \frac{\omega_A ^2}{\omega_\nu \omega_\eta},\; \frac{\pi^2 \textrm{Q}\textrm{Pm}}{a^4} \rightarrow \frac{\omega_A ^2}{ \omega_\eta ^2},\; \frac{\textrm{Ta}}{a^4} \rightarrow \frac{4\Omega^2}{\omega_\nu^2},\; \frac{\pi^2}{a^2}\rightarrow \alpha^2=\left(\frac{k_z}{|{\bf k}|}\right)^2\]
the expression for $\textrm{Ro}_{\textrm{cr}}$ is found  \cite{37s}:
\[\textrm{Ro}_{\textrm{cr}}=-\frac{{(\omega _A^2+\omega_\nu\omega_\eta)^2+4\alpha^2\Omega^2 \omega_\eta^2 }}{{4\Omega^2 \alpha^2 (\omega_A^2 +\omega_\eta^2 )}}  \]
According to the dispersion equation (\ref{eq40}) with $\textrm{Ra}=0$ and $\Omega=0$, an expression for the critical value of the magnetic Rossby number $\textrm{Rb}_{cr}$ is following:
\[\textrm{Rb}_{cr}=\frac{a^2(a^4+\pi^2\textrm{Q})^2-4\pi^4\textrm{Q}^2\xi^2}{4\pi^2\xi^2\textrm{Q}(a^4+\pi^2\textrm{Q})}, \]
or in dimensional variables \cite{37s}
\[ \textrm{Rb}_{cr}=\frac{(\omega_\nu \omega_\eta+\omega_A^2)^2-4\alpha^2\omega_A^2\omega_{A\phi}^2}{4\alpha^2\omega_{A\phi}^2(\omega_\nu \omega_\eta+\omega_A^2)}, \]
where notations for viscosity $\omega_{\nu} =\nu k^{2} $, ohmic $\omega_{\eta }=\eta k^{2}$ frequency, Alfven frequency  $\omega_{A}^{2} =k_{z}^{2} c_{A}^{2} =\frac{k_{z}^{2} B_{0}^{2} }{4\pi \rho _{0} }$, $\omega_{A\phi}^2=\frac{B_{0\phi}^2}{4\pi \rho_{0}R_0^2}$ are outlined.
Let us proceed to the study of a more general case, when there is a heating of a liquid layer $\textrm{Ra}\neq 0$  and its nonuniform rotation $\textrm{Ro}\neq 0 $ in a spiral magnetic field $\xi\neq 0$. Here we will consider a convective flow in a flat nonuniformly rotating layer in the form of cells. The magnitude of the growth rate of disturbances $\gamma$  in the general case is complex $\gamma=\gamma_r+i\omega_i$. It is obvious that the system is stable if $\gamma_r<0$, and unstable if $\gamma_r>0$. Let us proceed to the definition of the stability boundary for monotonic ($\omega_i=0$) and vibrational perturbations ($\omega_i\neq 0$). At the boundary of stability (neutral states),$\gamma_r=0$, therefore making the change in equation (\ref{eq40}) $\gamma=i\omega_i$ we found:
\begin{equation} \label{eq41} \textrm{Ra}=\textrm{Ra}_r+i\omega_i\textrm{Ra}_i,
\end{equation}
where
\[\textrm{Ra}_r=\frac{a^2}{k^2(a^4+\omega_i^2\textrm{Pm}^2)}\left[(a^4+\pi^2\textrm{Q}-\omega_i^2\textrm{Pm})(a^4+\omega_i^2\textrm{Pm}\Pr)-\omega_i^2 a^4(1+\textrm{Pm})(\Pr-\textrm{Pm}) \right]+   \]
\[+\pi^2 \textrm{Ta}(1+\textrm{Ro})\cdot \frac{(a^4-\omega_i^2\textrm{Pm}\Pr)(a^4+\pi^2\textrm{Q}-\omega_i^2\textrm{Pm})+\omega_i^2 a^4(1+\textrm{Pm})(\textrm{Pm}+\Pr) }{k^2((a^4+\pi^2\textrm{Q}-\omega_i^2\textrm{Pm})^2+\omega_i^2a^4(1+\textrm{Pm})^2)}+\]
\[+\pi^4\textrm{Q}(\textrm{TaRoPm}-4\textrm{Q}\xi^2)\cdot \frac{(a^4+\omega_i^2\textrm{Pm}\Pr)(a^4+\pi^2\textrm{Q}-\omega_i^2\textrm{Pm})+\omega_i^2 a^4(1+\textrm{Pm})(\Pr-\textrm{Pm})}{k^2(a^4+\omega_i^2\textrm{Pm}^2)((a^4+\pi^2\textrm{Q}-\omega_i^2\textrm{Pm})^2+\omega_i^2a^4(1+\textrm{Pm})^2)}-\]
\[-4\pi^2\xi^2\textrm{QRb}\cdot\frac{a^4+\omega_i^2\Pr\textrm{Pm}}{k^2(a^4+\omega_i^2\textrm{Pm}^2)},\]
\[\textrm{Ra}_i=\frac{a^4}{k^2(a^4+\omega_i^2\textrm{Pm}^2)}\left[(1+\textrm{Pm})(a^4+\omega_i^2\textrm{Pm}\Pr)+(\Pr-\textrm{Pm})(a^4+\pi^2\textrm{Q}-\omega_i^2\textrm{Pm})\right] + \]
\[+\pi^2 \textrm{Ta}(1+\textrm{Ro})\cdot \frac{a^2\left[(\Pr+\textrm{Pm})(a^4+\pi^2\textrm{Q}-\omega_i^2\textrm{Pm})-(1+\textrm{Pm})(a^4-\omega_i^2\textrm{Pm}\Pr)\right]}{k^2((a^4+\pi^2\textrm{Q}-\omega_i^2\textrm{Pm})^2+\omega_i^2a^4(1+\textrm{Pm})^2)}+\]
\[+\pi^4\textrm{Q}(\textrm{TaRoPm}-4\textrm{Q}\xi^2)\cdot \frac{a^2\left[(\Pr-\textrm{Pm})(a^4+\pi^2\textrm{Q}-\omega_i^2\textrm{Pm})-(1+\textrm{Pm})(a^4+\omega_i^2\textrm{Pm}\Pr)\right]}{k^2(a^4+\omega_i^2\textrm{Pm}^2)((a^4+\pi^2\textrm{Q}-\omega_i^2\textrm{Pm})^2+\omega_i^2a^4(1+\textrm{Pm})^2)}- \]
\[-4\pi^2\xi^2\textrm{QRb}\cdot\frac{a^2(\Pr-\textrm{Pm})}{k^2(a^4+\omega_i^2\textrm{Pm}^2)}. \]

\subsection{Stationary convection regime }

In the case when the coefficient  $\gamma$  is zero $(\gamma=0)$, then from formula (\ref{eq40}) we can find the critical value of the Rayleigh number Rast for stationary convection:
\begin{equation} \label{eq42}
  \textrm{Ra}_{st}  = \frac{a^6}{k^2} + \frac{\pi^2 a^2\textrm{Q}}{k^2} + \frac{\pi^2a^4 \textrm{Ta}}{k^2 (a^4 + \pi^2\textrm{Q})} + \frac{\pi^2  \textrm{TaRo}(a^4+\pi^2\textrm{QPm})-4\pi^4\xi^2 \textrm{Q}^2}{k^2(a^4+\pi^2\textrm{Q})}-\frac{4\pi^2}{k^2}\cdot\xi^2\textrm{QRb} \end{equation}
The minimum value of the critical Rayleigh number is found from the condition $\partial\textrm{Ra}_{st}/\partial k =0$ and corresponds to the wave numbers $k=k_c$ that satisfy the following equation:
\[\frac{{2k_c ^2-\pi^2 }}{{k_c }}-\frac{{\pi ^4\textrm{Q}}}{{k_c (\pi^2+k_c^2)^2 }}+\frac{{2\pi^2 k_c\textrm{Ta}(1+\textrm{Ro})}}{(\pi^2+k_c^2)\left( (\pi^2+k_c^2)^2+\pi^2\textrm{Q}\right)} - \]
\begin{figure}
  \centering
	\includegraphics[width=17 cm, height=16 cm]{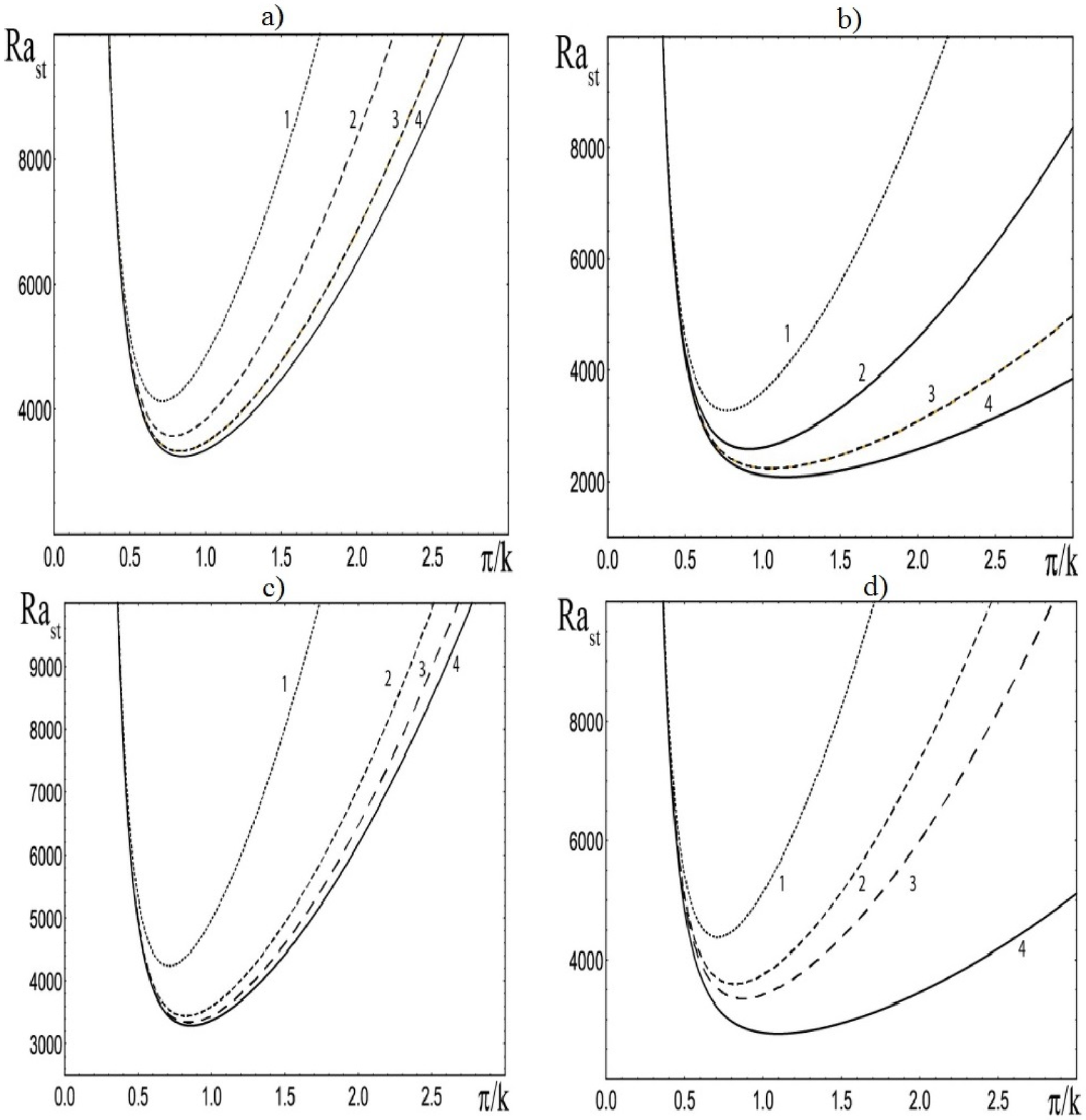}  \\
   \caption{ Dependence of the stationary Rayleigh number  $\textrm{R}_{st}$  on the wave numbers $\pi/k$ with constant parameters $\textrm{Q}=150$, $\textrm{Ta}=500$, $\textrm{Pm}=1$: (a)  $\xi=0$: curve 1  -- $\textrm{Ro}=2$, curve 2 -- $\textrm{Ro}=0$, curve 3 -- $\textrm{Ro}=-3/4$, curve 4 -- $\textrm{Ro}=-1$;  (b)  $\xi=1$ and $\textrm{Rb}=0$:  curve 1 -- $\textrm{Ro}=2$, curve 2 -- $\textrm{Ro}=0$, curve 3 -- $\textrm{Ro}=-3/4$, curve 4 -- $\textrm{Ro}=-1$;  (c)  $\xi=1$:  curve 1 -- $\textrm{Ro}=2$ and $\textrm{Rb}=-1$, curve 2 -- $\textrm{Ro}=-3/4$ and $\textrm{Rb}=-1$, curve 3 -- $\textrm{Ro}=-1$ and $\textrm{Rb}=-1$, curve 4 -- $\textrm{Ro}=0 $  and $\textrm{Rb}=0$; (d)  $\xi=1.5$:  curve 1 -- $\textrm{Ro}=2$ and $\textrm{Rb}=-1$, curve 2 -- $\textrm{Ro}=-3/4$ and $\textrm{Rb}=-1$, curve 3 -- $\textrm{Ro}=-1$ and $\textrm{Rb}=-1$, curve 4 -- $\textrm{Ro}=0 $  and $\textrm{Rb}=0$.}\label{fg4}
   \end{figure}
\[-\frac{\pi^2 \textrm{Ta}((\pi^2+k_c^2)^2+\pi^2\textrm{Q}+2k_c^2(\pi^2+ k_c^2))}{k_c((\pi^2+k_c^2)^2  + \pi ^2\textrm{Q})^2}-\]

\[-\frac{\pi ^2 \textrm{TaRo}((\pi^2  + k_c^2 )^2  + \pi ^2\textrm{QPm})((\pi^2+ k_c^2 )^2  + \pi ^2\textrm{Q}+2k_c^2(\pi ^2  + k_c ^2))}{k_c(\pi ^2  + k_c^2)^2((\pi^2+k_c^2)^2+\pi^2\textrm{Q})^2}+\]

\[+4\pi^4\xi^2\textrm{Q}^2\cdot\frac{2((\pi^2+ k_c^2)^2+\pi^2\textrm{Q})+4k_c^4(\pi^2+ k_c^2)}{k_c^3((\pi^2+ k_c^2)^2+\pi^2\textrm{Q})^2}+4\pi^2\xi^2\textrm{QRb}\cdot\frac{2}{k_c^3}=0 \]
Graphs in Fig. \ref{fg4} show the dependence of the stationary Rayleigh number $ \textrm{Ra}_{st}$  on the wave numbers $\pi/k$ for various profiles of non-uniform rotation $( \textrm{Ro} )$ and magnetic field  $( \textrm{Rb} )$. The minimum number of $\textrm{Ra}_{st}^{\textrm{min}}$ in Fig. \ref{fg4} corresponds to a point on the neutral curve separating the regions of stable and unstable perturbations. Fig. \ref{fg4}$\textrm{a}$ shows the case when the azimuthal magnetic field is not $\xi=0$. Here we see that with an increase in the positive profile of the Rossby number $\textrm{Ro}$ the minimum value of the stationary Rayleigh number $\textrm{Ra}_{st}^{\textrm{min}}$ also increases, i.e. raises the threshold for the growth of instability. On the other hand, for negative rotation profiles: Keplerian $( \textrm{Ro}=-3/4 )$ and Rayleigh $( \textrm{Ro}=-1 )$, we observe a decrease in the critical Rayleigh number, i.e. a lower threshold for the growth of instability compared with the case of homogeneous $( \textrm{Ro}=0 )$   and nonuniform $( \textrm{Ro}=2 )$   rotation. These results correspond to \cite{39s}. The case of a homogeneous $( \textrm{Rb}=0 )$  azimuthal magnetic field with the parameter $\xi=1$  is shown in Fig. \ref{fg4}$\textrm{b}$. This shows that a uniform azimuthal magnetic field contributes to lowering the instability threshold, i.e. the minimum critical Rayleigh numbers for different rotation profiles (Rossby numbers  $ \textrm{Ro}$) are smaller than at  $\xi=0$  (see Fig. \ref{fg4}$\textrm{a}$). In the case of $\xi=1$ and $\textrm{Rb}=0$, we see that the negative rotation profiles $\textrm{Ro}<0$ (curves 3,4) have critical Rayleigh numbers that are smaller in magnitude than for the positive rotation profiles  $\textrm{Ro}>0$  (curves 1,2). The case of influence on the convective instability of an inhomogeneous azimuthal magnetic field $(\textrm{Rb}=-1 )$  with the parameter $\xi=1$ is shown in Fig. \ref{fg4}$\textrm{c}$. Compared to the uniform (or solid body) case $\textrm{Ro}=0 $ and  $\textrm{Rb}=0$ (curve 4), we see an increase in the instability threshold (an increase in the minimum stationary Rayleigh number $\textrm{Ra}_{st}^{\textrm{min}}$) towards positive Rossby $\textrm{Ro}$ numbers (see curves 1, 2, 3). A similar picture (see Fig. \ref{fg4}$\textrm{d}$) is observed with the parameter $\xi=1.5$ and a nonuniform $( \textrm{Rb}=-1 )$ azimuthal magnetic field.

To study the effects of nonuniform rotation and spiral magnetic field, we have calculated the derivatives $d \widetilde R/d \widetilde Q$, $d \widetilde R/d \widetilde T$, $d \widetilde R/d \textrm{Ro}$, $d \widetilde R/d \widetilde\xi$, $d \widetilde R/d \textrm{Rb}$ in new variables
\[\widetilde R=\frac{\textrm{Ra}_{st}}{\pi^4},\quad \widetilde T=\frac{\textrm{Ta}}{\pi^4}, \quad \widetilde Q=\frac{\textrm{Q}}{\pi^2}, \quad \textrm{x}=\frac{k^2}{\pi^2}, \quad \widetilde \xi= \frac{\xi}{\pi}\]
which have the form
\begin{equation}\label{eq42m}
\frac{d \widetilde R}{d \widetilde Q}=\frac{1+\textrm{x}}{\textrm{x}}-\frac{(1+\textrm{x})^2\widetilde T(\textrm{Ro}(\textrm{Pm}-1)-1) }{\textrm{x}((1+\textrm{x})^2+\widetilde Q)^2}-\frac{4\widetilde \xi^2}{\textrm{x}}\cdot \left(\textrm{Rb}+\frac{\widetilde Q(2(1+\textrm{x})^2+\widetilde Q)}{(1+\textrm{x})^2+\widetilde Q)^2}\right) ,
 \end{equation}
\begin{equation}\label{eq42n}
\frac{d \widetilde R}{d \widetilde T}=\frac{(1+\textrm{x})^2}{\textrm{x}((1+\textrm{x})^2+\widetilde Q)}+\frac{\textrm{Ro}((1+\textrm{x})^2+\widetilde Q \textrm{Pm})}{\textrm{x} ((1+\textrm{x})^2+\widetilde Q)} ,
\end{equation}
\begin{equation}\label{eq42p}
\frac{d \widetilde R}{d \textrm{Ro}}=\frac{\widetilde T ((1+\textrm{x})^2+\widetilde Q \textrm{Pm}) }{\textrm{x}((1+\textrm{x})^2+\widetilde Q)} ,
\end{equation}
\begin{equation}\label{eq42r}
\frac{d \widetilde R}{d \widetilde\xi}=-\frac{8\widetilde\xi \widetilde Q }{\textrm{x}}\cdot \frac{\widetilde Q(1+\textrm{Rb})-\textrm{Rb}(1+\textrm{x})^2 }{(1+\textrm{x})^2+\widetilde Q}, \quad \frac{d \widetilde R}{d \textrm{Rb}}=-\frac{4\widetilde \xi^2 \widetilde Q}{\textrm{x}} .
\end{equation}
Formula (\ref{eq42m})  shows that $d \widetilde R/d \widetilde Q$ can be positive or negative, i.e. the axial magnetic field (Chendrasekar number $\widetilde Q$) has a stabilizing or destabilizing effect on stationary convection. The destabilizing effect occurs when the conditions for profiles of nonuniform rotation (Rossby number  $\textrm{Ro}$) and nonuniform azimuthal magnetic field (magnetic Rossby number  $\textrm{Rb}$) are met:
\[ \textrm{Ro}(\textrm{Pm}-1)>1, \; \textrm{Rb}>\frac{\widetilde Q(2(1+\textrm{x})^2+\widetilde Q)}{(1+\textrm{x})^2+\widetilde Q)^2} \,(\textrm{when}~ \textrm{Rb}>0),\]
\[ \textrm{Rb}<\frac{\widetilde Q(2(1+\textrm{x})^2+\widetilde Q)}{(1+\textrm{x})^2+\widetilde Q)^2} \,(\textrm{when}~ \textrm{Rb}<0).   \] 
The first inequality holds for positive Rossby numbers $\textrm{Ro}>0$  and Prandtl magnetic numbers  $\textrm{Pm}>1$, or $\textrm{Ro}<0$ and $\textrm{Pm}<1$. It follows from formulas (\ref{eq42n})-(\ref{eq42p}) that nonuniform rotation has a stabilizing effect ($d \widetilde R/d \widetilde T>0$, $d \widetilde R/d \textrm{Ro}>0$ )  in the case
\begin{figure}
  \centering
	\includegraphics[width=17 cm, height=16 cm]{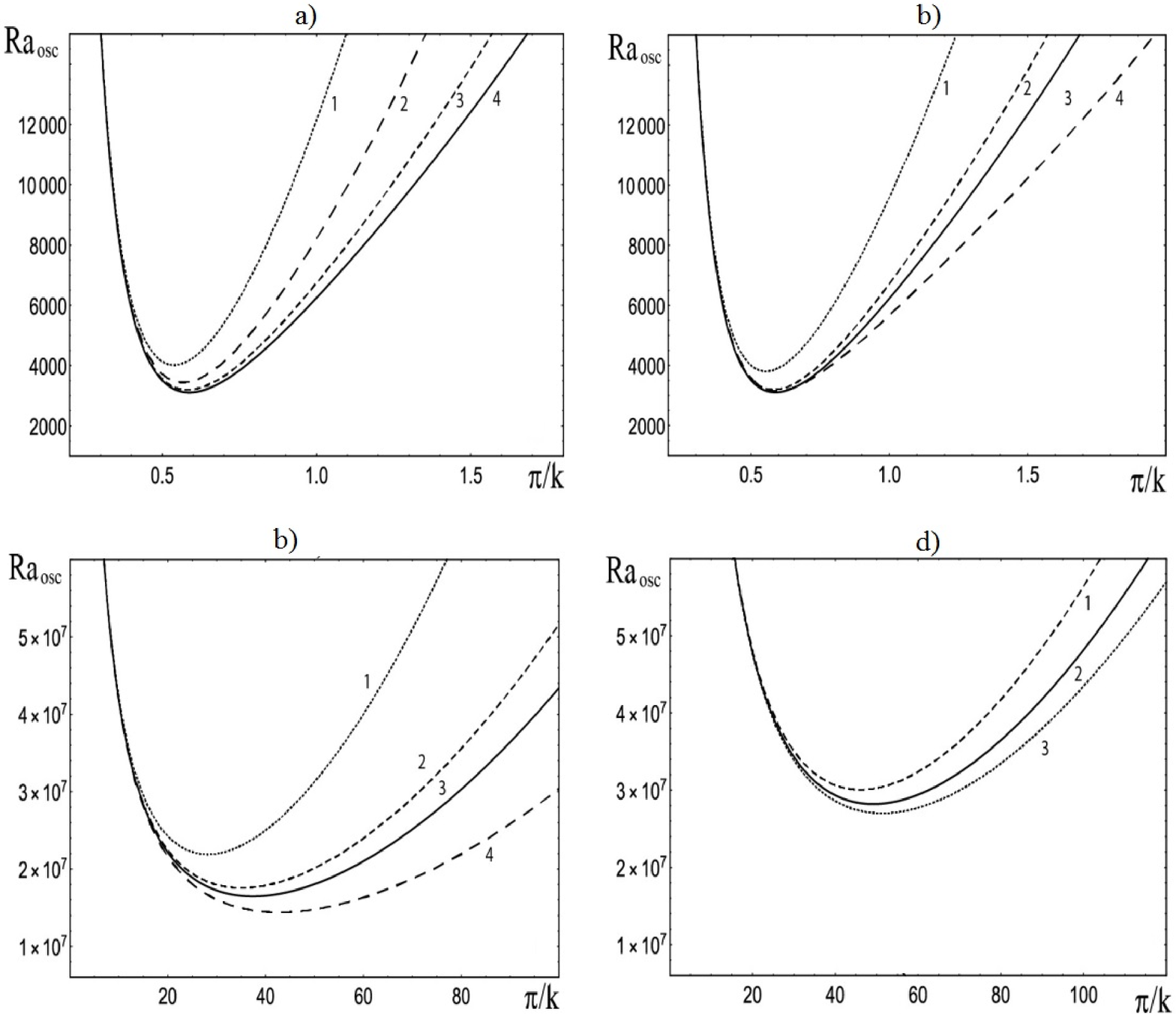}  \\
   \caption{ Neutral stability curves $(\pi/k,\textrm{Ra}_{\textrm{osc}})$ of oscillatory convection with constant parameters  $\textrm{Q}=150$, $\textrm{Ta}=500$, $\textrm{Pm}=1$:   (a)  $\xi=0$:  curve 1 -- $\textrm{Ro}=2$, curve 2 -- $\textrm{Ro}=0$, curve 3 -- $\textrm{Ro}=-3/4$, curve 4 -- $\textrm{Ro}=-1$; (b)  $\xi=1$ and $\textrm{Rb}=0$:  curve 1 -- $\textrm{Ro}=2$, curve 2 -- $\textrm{Ro}=-3/4$, curve 3 -- $\textrm{Ro}=-1$, curve 4 -- $\textrm{Ro}=0$;  (c)  $\xi=1$ and $\textrm{Rb}=-1$:   curve 1 -- $\textrm{Ro}=2$, curve 2 -- $\textrm{Ro}=-3/4$, curve 3 -- $\textrm{Ro}=-1$, curve 4 -- $\textrm{Ro}=0 $; (d)  $\xi=1.5$  and $\textrm{Rb}=-1$:  curve 1 -- $\textrm{Ro}=-3/4$, curve 2 -- $\textrm{Ro}=-1$, curve 3 -- $\textrm{Ro}=2$. }\label{fg5}
   \end{figure}
of positive Rossby numbers $\textrm{Ro}>0$, and otherwise for $\textrm{Ro}<0$ nonuniform rotation can have a destabilizing effect ($d \widetilde R/d \widetilde T<0$, $d \widetilde R/d \textrm{Ro}<0$). Formulas (\ref{eq42r}) show that a uniform azimuthal magnetic field ($\textrm{Rb}=0$) has a destabilizing effect ($d \widetilde R/d \widetilde\xi<0$)  on stationary convection. In contrast, a nonuniform azimuthal magnetic field with a negative profile $\textrm{Rb}<0$ has a stabilizing ($d \widetilde R/d \textrm{Rb} >0$)  effect.

The expression for the stationary Rayleigh number (\ref{eq42}) in the limiting case of a uniform axial magnetic field $\xi=0$ coincides with the result of \cite{42s}, and the limit of uniform rotation is  $\textrm{Ro}=0$ and $\xi=0$ with the result of Chandrasekhar \cite{1s}.

\subsection{Vibrational (oscillating) convection regime}

Obviously, for physical reasons, the value of $\textrm{Ra}$ is real, then the imaginary part in (\ref{eq41}) should be zero. The following options are possible $\omega_i=0$  and $\textrm{Ra}_i=0 $. In the first variant ($\omega_i=0$), we have obtained the critical value of the Rayleigh number $\textrm{Ra}_{c}$ for monotone perturbations, which coincides with expression  (\ref{eq42}) for the stationary convection mode: $\textrm{Ra}_{c}=\textrm{Ra}_{st}$. In the case of an oscillatory perturbation $\omega_i \neq 0$  $(\textrm{Ra}_i=0)$, we find the critical Rayleigh number for oscillatory instability using the formula (\ref{eq29}):
\begin{equation} \label{eq43}	
\textrm{Ra}_{\textrm{osc}}=\frac{a^2}{k^2(a^4+\omega^2\textrm{Pm}^2)}\left[(a^4+\pi^2\textrm{Q}-\omega^2\textrm{Pm})(a^4+\omega^2\textrm{Pm}\Pr)-\omega^2 a^4(1+\textrm{Pm})(\Pr-\textrm{Pm}) \right]+ $$
$$+\pi^2 \textrm{Ta}(1+\textrm{Ro})\cdot \frac{(a^4-\omega^2\textrm{Pm}\Pr)(a^4+\pi^2\textrm{Q}-\omega^2\textrm{Pm})+\omega^2 a^4(1+\textrm{Pm})(\textrm{Pm}+\Pr) }{k^2((a^4+\pi^2\textrm{Q}-\omega^2\textrm{Pm})^2+\omega^2a^4(1+\textrm{Pm})^2)}+$$
$$+\pi^4\textrm{Q}(\textrm{TaRoPm}-4\textrm{Q}\xi^2)\cdot \frac{(a^4+\omega^2\textrm{Pm}\Pr)(a^4+\pi^2\textrm{Q}-\omega^2\textrm{Pm})+\omega^2 a^4(1+\textrm{Pm})(\Pr-\textrm{Pm})}{k^2(a^4+\omega^2\textrm{Pm}^2)((a^4+\pi^2\textrm{Q}-\omega^2\textrm{Pm})^2+\omega^2a^4(1+\textrm{Pm})^2)}-$$
$$-4\pi^2\xi^2\textrm{QRb}\cdot\frac{a^4+\omega^2\Pr\textrm{Pm}}{k^2(a^4+\omega^2\textrm{Pm}^2)}, \end{equation}
and the frequency of neutral vibrations $\omega=\omega_i$, satisfying the following equation:
\[\omega^6\cdot K_0 +\omega^4\cdot K_1 +\omega^2 \cdot K_2+K_3=0, \]
where
\[K_0=\textrm{Pm}^4(1+\Pr),\]
\[K_1= a^4(1+\textrm{Pr})\textrm{Pm}^2+\textrm{Pm}^2(\Pr-\textrm{Pm})\pi^2\textrm{Q}+\textrm{Pm}^2(1+\Pr)(a^4(1+\textrm{Pm})^2-2\textrm{Pm}(a^4+\pi^2\textrm{Q}))+\]
\[+\frac{\pi^2}{a^2}\textrm{Ta}(1+\textrm{Ro})\textrm{Pm}^4(\Pr-1)-\frac{4\pi^2\xi^2}{a^2}\textrm{QRb}(\Pr-\textrm{Pm})\textrm{Pm}^2 ,\]
\[K_2=(a^4(1+\textrm{Pr})+(\Pr-\textrm{Pm})\pi^2\textrm{Q})(a^4(1+\textrm{Pm})^2-2\textrm{Pm}(a^4+\pi^2\textrm{Q}))+\textrm{Pm}^2(1+\Pr)(a^4+\pi^2\textrm{Q})^2+  \]
\[+\frac{\pi^2}{a^2}\textrm{Ta}(1+\textrm{Ro})\left(2a^4\textrm{Pm}^2(\Pr-1)+\pi^2\textrm{Q}(\Pr+\textrm{Pm})\textrm{Pm}^2 \right)- \]
\[-\frac{\pi^4 \textrm{Q}}{a^2}\left(\textrm{TaRoPm}-4 \textrm{Q}\xi^2\right)\left(2\textrm{Pm}\Pr+\textrm{Pm}^2(\Pr-1)\right)-\]
\[-\frac{4\pi^2\xi^2}{a^2}\left(\Pr-\textrm{Pm}\right)(\Pr-\textrm{Pm})\left(a^4(1+\textrm{Pm})^2-2\textrm{Pm}(a^4+\pi^2\textrm{Q})\right), \]
\[K_3=(a^4(1+\textrm{Pr})+(\Pr-\textrm{Pm})\pi^2\textrm{Q})(a^4+\pi^2\textrm{Q})^2+\frac{\pi^2}{a^2}\textrm{Ta}(1+\textrm{Ro})a^4 \times $$
\begin{figure}
  \centering
	\includegraphics[width=8 cm, height=5 cm]{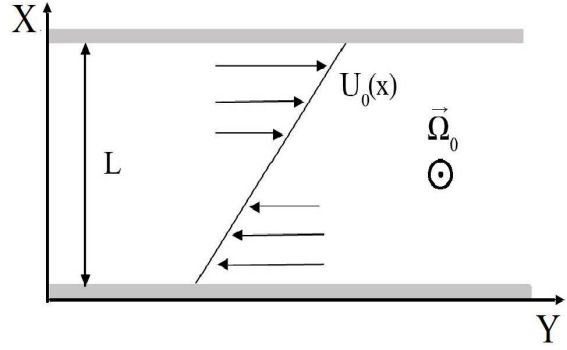}   \\ 
\caption{Scheme of the shear flow in rotating flows, the flow being approximated in the local Cartesian coordinate system as a linear shift
with velocity ${\bf{U}}_0(X)$, with the value of the flow being limited by coordinates  $X \in [0,L]$ and $Z \in [0,h]$, and by coordinate
$Y$ being unlimited. }\label{fg6}
\end{figure}
$$\times \left((\Pr+\textrm{Pm})\pi^2\textrm{Q}+a^4(\Pr-1)\right)+\frac{\pi^4 \textrm{Q}}{a^2}(\textrm{TaRoPm}-4\textrm{Q}\xi^2)(a^4(\Pr-2\textrm{Pm}-1)+ \pi^2\textrm{Q}(\Pr-\textrm{Pm}))-\]
\[- \frac{4\pi^2\xi^2}{a^2}\textrm{QRb}(\Pr-\textrm{Pm})(a^4+\pi^2\textrm{Q})^2. \]
Formula (\ref{eq43}), in some limiting cases, contains known results. For example, for the case of uniform rotation $(\textrm{Ro}=0)$ and non-conducting fluid $(\sigma=0)$, Chandrasekhar \cite{1s} obtained the expression for the critical Rayleigh number of oscillatory instability:
\[\textrm{Ra}_{\textrm{osc}}=\frac{1}{k^2}\left[a^6-\omega^2a^2\Pr+\frac{\pi^2\textrm{Ta}(a^4+\omega^2\Pr)}{a^4+\omega^2}\right] \]
In the absence of rotation $(\textrm{Ta}=0, \textrm {Ro}=0 )$, in a magnetized conducting fluid $\textrm{Q}\neq 0$ , the critical Rayleigh number $\textrm{Ra}_{\textrm{osc}}$  for vibrational convection was also obtained by Chandrasekhar \cite{1s}:
\[ \textrm{Ra}_{\textrm{osc}}=\frac{a^2}{k^2} \left[a^4-\omega^2\Pr+\frac{\pi^2 \textrm{Q}(a^4+\omega^2\Pr\textrm{Pm})}{a^4+\omega^2\textrm{Pm}^2}\right] \]
Fig. \ref{fg5} shows the dependence of the critical Rayleigh number $\textrm{Ra}_{\textrm{osc}}$ for oscillatory instability on the wave numbers $\pi/k$ for different profiles of nonuniform rotation $\textrm{Ro}$, nonuniform magnetic field $\textrm{Rb}$ and various values of the parameter $\xi$. On the graph Fig. \ref{fg5}$\textrm{a}$ shows the case when the azimuthal magnetic field is not $\xi=0$, which corresponds to the results of \cite{39s}. From which it follows that with negative Rossby numbers $\textrm{Ro}<0$ the threshold Rayleigh number $\textrm{Ra}_{\textrm{min}}^{\textrm{osc}}$ decreases. The case of a uniform azimuthal magnetic field ($\textrm{Rb}=0$)  with the parameter $\xi=1$ is shown in Fig. \ref{fg5}$\textrm{b}$ . Here the minimum threshold Rayleigh number for oscillating convection $\textrm{Ra}_{\textrm{min}}^{\textrm{osc}}$ corresponds to the profile for uniform rotation $\textrm{Ro}=0$. A similar situation arises in the case of a nonuniform azimuthal magnetic field ($\textrm{Rb}=-1$) with parameters $\xi=1$ (see Fig. \ref{fg5}$\textrm{c}$) and $\xi=1.5$ (see Fig. \ref{fg5}$\textrm{d}$). Here it can be seen that the minimum Rayleigh numbers  $\textrm{Ra}_{\textrm{min}}^{\textrm{osc}}$ are significantly larger than the cases shown in Fig. \ref{fg5}$\textrm{a}$-\ref{fg5}$\textrm{b}$.
Consequently, the vibrational instability thresholds for convection in a spiral magnetic field are higher than for convection in a constant axial magnetic field, regardless of the profile of nonuniform rotation (Rossby numbers $\textrm{Ro}$).

\section{Weakly nonlinear convection regime}

It is convenient to switch from a cylindrical coordinate system $(R,\varphi,z)$ to a local Cartesian $(X,Y,Z)$ to describe nonlinear convective phenomena in a nonuniformly rotating layer of an electrically conducting fluid. If we consider a fixed region of a fluid layer with a radius $R_0$ and an angular velocity of rotation $\Omega_0=\Omega(R_0)$, then the coordinates $X=R-R_0$  correspond to the radial direction, $Y=R_0(\varphi-\varphi_0)$ to the azimuth and $Z=z$ -- to the vertical direction (see Fig. \ref{fg6}). Then, nonuniform  rotation of the fluid layer can be represented locally as a rotation with a constant angular velocity  ${\bf{\Omega}}_0$   and azimuthal width \cite{40s}, whose velocity profile is locally linear: ${\bf{U}}_0=-q {\Omega}_0 X {\bf{e}}_y$, where $q\equiv -d \ln \Omega/d\ln R$  is a dimensionless width parameter defined by from the profile of angular velocity of rotation $\Omega (R)=\Omega_0 (R/R_0)^{-q}$. The global parameter $q$ is related to the hydrodynamic Rossby number $\textrm{Ro}=\frac{R}{2\Omega} \frac{\partial\Omega}{\partial R}$ by the relation: $q=-2\textrm{Ro}$. Note that the accretion disks with the spherical parameter $q=3/2$ $(\textrm{Ro}=-3/4)$ correspond to the Keplerian disk, $q=2$ $(\textrm{Ro}=-1) $  corresponds to the disk with a constant angular momentum or Rayleigh rotation profile. The case $q=1$ $(\textrm{Ro}=-1/2)$  corresponds to a system with a flat rotation curve, while $q=0$ $(\textrm{Ro}=0)$  corresponds to a uniform (or solid-body) rotation with a constant angular velocity.
When moving from cylindrical to local Cartesian coordinate system, the equations for perturbed quantities $({\bf{u}}=(u_X,u_Y,u_Z), {\bf{b}}=(b_X,b_Y,b_Z), p, \theta)$  will take the following form:
\begin{equation} \label{eq44} 
\left(\frac{\partial}{\partial t}-\nu\nabla^2\right)u_X+({\bf {u}} \nabla)u_X-2\Omega_0u_Y=-\frac{1}{\rho_0}\frac{\partial \widetilde p}{\partial X}-\textrm{s} b_Y+\frac{1}{4\pi \rho_0}({\bf {b}} \nabla)b_X+\frac{B_{0z}}{4\pi \rho_0}\frac{\partial b_X}{\partial Z}
\end{equation}
\begin{equation} \label{eq45} 
\left(\frac{\partial}{\partial t}-\nu\nabla^2\right)u_Y+({\bf {u}} \nabla)u_Y +2\Omega_0u_X(1+\textrm{Ro})= \textrm{s}(1+\textrm{Rb})b_X + \frac{1}{4\pi \rho_0}({\bf {b}} \nabla)b_Y+ \frac{B_{0z}}{4\pi \rho_0}\frac{ \partial b_Y }{\partial Z}
\end{equation}
\begin{equation} \label{eq46} 
\left(\frac{\partial}{\partial t}-\nu\nabla^2\right)u_Z+({\bf {u}}\nabla)u_Z=-\frac{1}{\rho_0}\frac{\partial \widetilde p}{\partial Z}+g\beta\theta+\frac{1}{4\pi \rho_0}({\bf {b}} \nabla)b_Z+\frac{B_{0z}}{4\pi\rho_0}\frac{\partial b_Z}{\partial Z}
\end{equation}
\begin{equation} \label{eq47}
\left(\frac{\partial}{\partial t}-\eta\nabla^2\right)b_X-B_{0z}\frac{\partial u_X}{\partial Z}+({\bf {u}} \nabla)b_X-({\bf {b}} \nabla)u_X=0
\end{equation}
\begin{equation} \label{eq48}
\left(\frac{\partial}{\partial t}-\eta\nabla^2\right)b_Y-B_{0z}\frac{\partial u_Y}{\partial Z}-2\Omega_0\textrm{Ro}b_X +({\bf {u}} \nabla)b_Y-({\bf {b}} \nabla)u_Y=0
\end{equation}
\begin{equation} \label{eq49}
\left(\frac{\partial}{\partial t}-\eta\nabla^2\right)b_Z-B_{0z}\frac{\partial u_Z}{\partial Z}+({\bf {u}}\nabla)b_Z-({\bf {b}} \nabla)u_Z=0
\end{equation}
\begin{equation} \label{eq50}
\left(\frac{\partial}{\partial t}-\chi\nabla^2\right)\theta-w \cdot \frac{T_d-T_u}{h} +({\bf {u}}\nabla)\theta=0, 
\end{equation}
where $\textrm{s}= B_{0\phi}(R_0)/2\pi\rho_0R_0$, the pressure $\widetilde{p}$ includes the disturbed magnetic pressure $p_m$: $\widetilde{p}=p+p_m$. The nabla operators can be described as:
$$({\bf a} \nabla)=a_x\,\frac{\partial}{\partial X}+a_z\,\frac{\partial}{\partial Z},\quad \nabla^2=\frac{\partial^2}{\partial X^2}+\frac{\partial^2}{\partial Z^2}. $$
In equations (\ref{eq44})-(\ref{eq50}) , we have considered that all perturbed quantities depend only on two variables $(X,Z)$, i.e. we consider the dynamics of axisymmetric perturbations. The solenoidal equations for axisymmetric velocity and magnetic field perturbations will take the form
\begin{equation} \label{eq51} \frac{\partial u_X}{\partial X}+\frac{\partial u_Z}{\partial Z}=0, \quad \frac{\partial b_X}{\partial X}+\frac{\partial b_Z}{\partial Z}=0
\end{equation}	
\begin{figure}
  \centering
	\includegraphics[width=18 cm, height=21 cm]{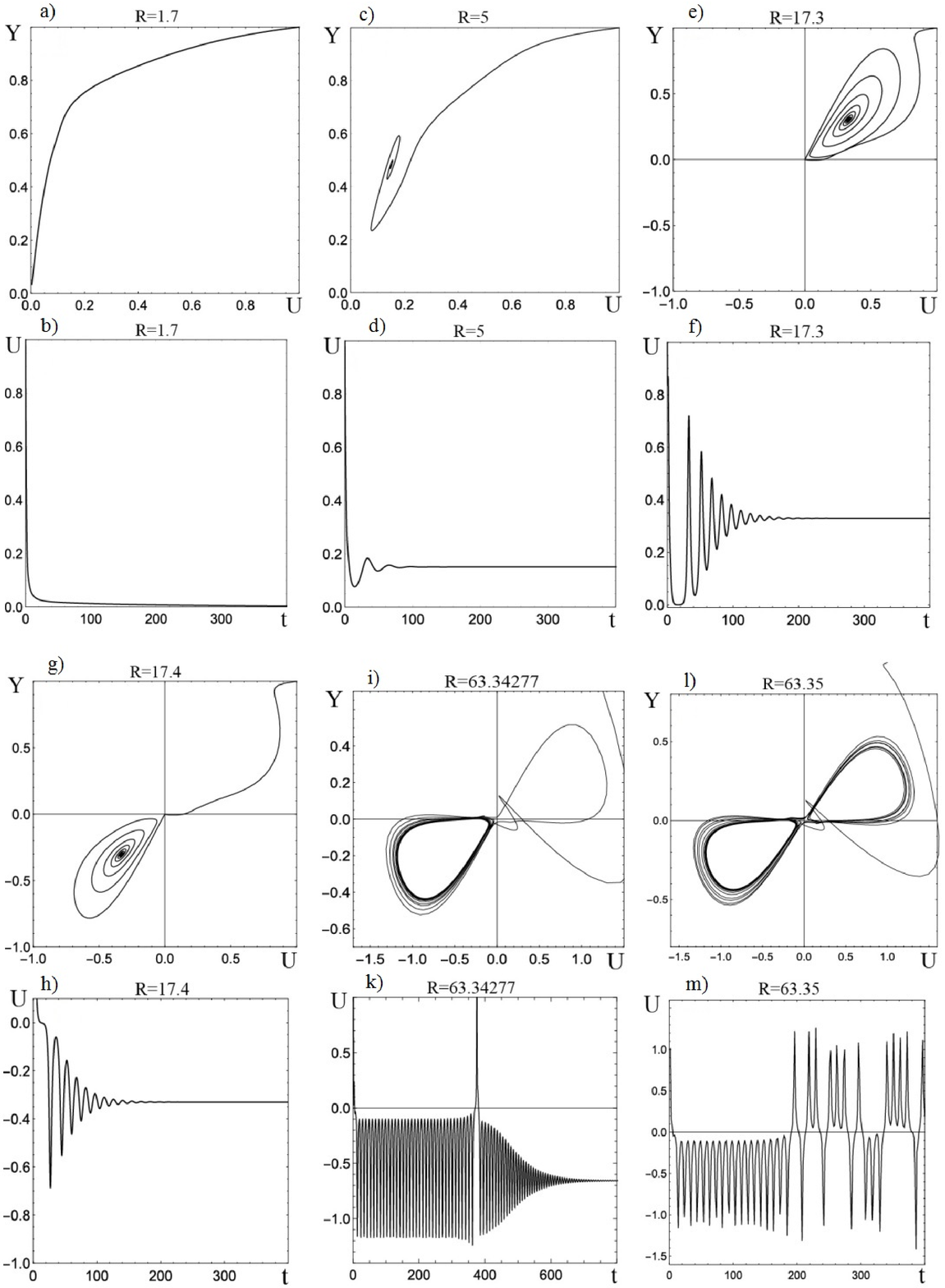} \\
\caption{ $\textrm{ a), c), e), g), i), l)}$ -- projections of phase trajectories in the $U-Y$ plane upon the variation of parameter $\textrm{R}$ for $\textrm{Ro}=-1, \textrm{Rb}=-1$; $\textrm{b), d), f), h), k), m)}$ is the time dependence of the amplitude variations of the magnetic component $U(\widetilde t)$. }
\label{fg7}
\end{figure} 
\begin{figure}
  \centering
	\includegraphics[width=18 cm, height=21 cm]{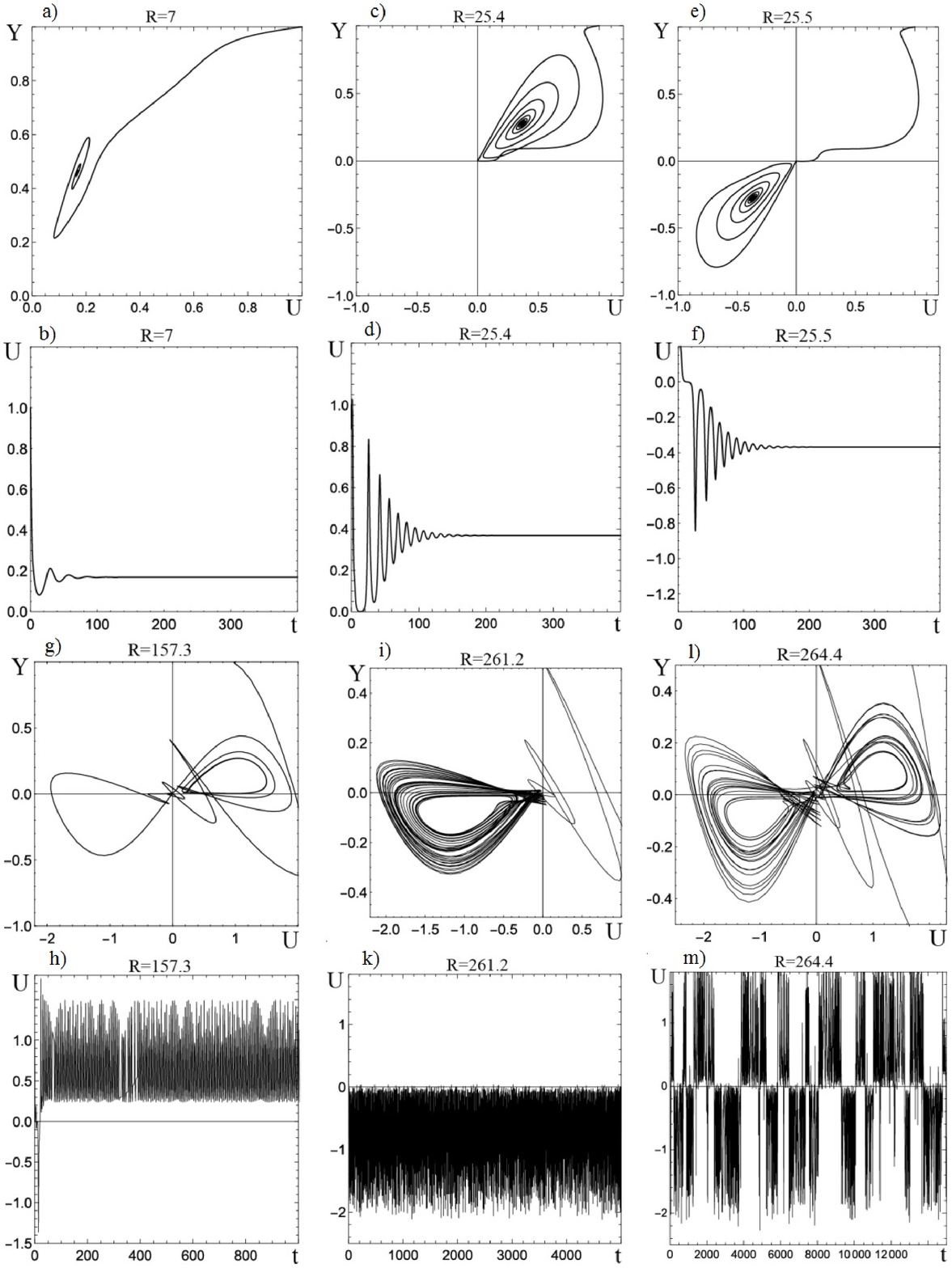} \\
\caption{ $\textrm{ a), c), e), g), i), l)}$ -- projections of phase trajectories in the $U-Y$ plane upon the variation of parameter $\textrm{R}$ for  $\textrm{Ro}=-3/4,  \textrm{Rb}=-1$; $\textrm{b), d), f), h), k), m)}$ -- is the time dependence of the amplitude variations of the magnetic component $U(\widetilde t)$.}\label{fg8}
\end{figure} 
According to equations (\ref{eq51}), we can introduce two scalar functions: the hydrodynamic function of the current $\psi$   and the magnetic function $\phi$, for which the following relations hold:
\[ u_X=-\frac{\partial\psi}{\partial Z},\quad u_Z=\frac{\partial\psi}{\partial X},\quad b_X=-\frac{\partial\phi}{\partial Z},\quad b_Z=\frac{\partial\phi}{\partial X}.  \]
Equations (\ref{eq44})-(\ref{eq50}) are written through the current functions $\psi$ and $\phi$, transferring the nonlinear terms to the right side of the equations:
\begin{equation} \label{eq52} 
\left(\frac{\partial }{{\partial t}}-\nu \nabla^2\right)\nabla^2 \psi+2\Omega_0 \frac{{\partial u_Y}}{{\partial Z}}-\frac{{B_{0z}}}{{4\pi \rho_0 }}\frac{\partial }{{\partial Z}}\nabla ^2 \phi-g\beta \frac{{\partial \theta }}{{\partial X}}-\textrm{s}\frac{\partial b_Y}{\partial Z}= \frac{1}{{4\pi \rho _0 }}J(\phi,\nabla^2 \phi ) - J(\psi,\nabla^2 \psi ) \end{equation}
\begin{equation} \label{eq53} 
\left(\frac{{\partial }}{{\partial t}}-\nu \nabla^2 \right) u_Y - 2\Omega_0 (1+\textrm{Ro})\frac{{\partial \psi }}{{\partial Z}}-\frac{{B_{0z} }}{{4\pi \rho_0 }}\frac{{\partial u_Y}}{{\partial Z}}-\textrm{s}(1+\textrm{Rb})b_X=\frac{1}{{4\pi \rho_0 }}J(\phi, b_Y) - J(\psi ,u_Y)
 \end{equation}
\begin{equation} \label{eq54}
\left(\frac{{\partial }}{{\partial t}}-\eta \nabla^2 \right)\phi - B_{0z} \frac{{\partial \psi }}{{\partial Z}}= - J(\psi,\phi )\end{equation}
\begin{equation} \label{eq55}
\left(\frac{{\partial }}{{\partial t}}-\eta \nabla^2\right) b_Y - B_{0z} \frac{{\partial u_Y}}{{\partial Z}} + 2\Omega_0 \textrm{Ro}\frac{{\partial \phi }}{{\partial Z}} -4\pi\rho_0\textrm{s}\textrm{Rb}\frac{\partial \psi}{\partial Z} = J(\phi,u_Y) - J(\psi,b_Y)\end{equation}
\begin{equation} \label{eq56}
\left(\frac{{\partial }}{{\partial t}}- \chi \nabla^2 \right) \theta- \frac{T_d-T_u}{h} \cdot \frac{{\partial \psi }}{{\partial x}} = - J(\psi,\theta )
\end{equation}
where
$$J(a,b) = \frac{{\partial a}}{{\partial x}}\frac{{\partial b}}{{\partial z}} - \frac{{\partial a}}{{\partial z}}\frac{{\partial b}}{{\partial x}}$$ 
-- Jacobian operator or Poisson bracket $J(a,b)\equiv \left\{a,b \right\}$.
In the absence of thermal phenomena and an external azimuth magnetic field $(\textrm{s}=0)$, the system of equations (\ref{eq52})-(\ref{eq56}) was used to study the saturation mechanism of the standard MRI \cite{41s}. Since here we take into account thermal phenomena, it is convenient to use dimensionless variables in equations (\ref{eq52})-(\ref{eq56}):
\[ (X,Z)=h(x^*,z^*), \; t=\frac{h^2}{\nu}t^*,\; \psi=\chi\psi^*,\; \phi=hB_0 \phi^*,\] 
\[ u_Y=\frac{\chi}{h}v^*,\; b_Y=B_0 \widetilde v^*,\; \theta=(T_d-T_u)\theta^*, B_0=B_{0z}. \]  
Omitting the asterisk symbol, we will rewrite equations  (\ref{eq52})-(\ref{eq56}) in dimensionless variables:
\begin{equation} \label{eq57}
\left(\frac{\partial }{{\partial t}}- \nabla ^2 \right)\nabla ^2 \psi  + \sqrt {\textrm{Ta}} \frac{{\partial v}}{{\partial z}} - \Pr\textrm{Pm}^{-1}\textrm{ Q}\frac{\partial }{{\partial z}}\nabla ^2 \phi  -\textrm{Ra}\frac{{\partial \theta }}{{\partial x}}-2\xi\Pr\textrm{Pm}^{-1}\textrm{ Q}\frac{\partial \widetilde v}{\partial z}=$$
$$=\Pr\textrm{Pm}^{-1}\textrm{Q}\cdot J(\phi,\nabla^2 \phi ) - {\Pr}^{-1}\cdot J(\psi,\nabla^2 \psi )\end{equation}	
\begin{equation} \label{eq58}
\left(\frac{{\partial }}{{\partial t}} -\nabla ^2 \right) v - \sqrt {\textrm{Ta}}(1 +\textrm{Ro})\frac{{\partial \psi }}{{\partial z}} - \Pr\textrm{Pm}^{ - 1}\textrm{Q}\frac{{\partial \widetilde v}}{{\partial z}}+2\xi\Pr\textrm{Pm}^{-1}\textrm{Q}(1+\textrm{Rb})\frac{\partial \phi}{\partial z} = $$
$$= \Pr\textrm{Pm}^{-1}\textrm{Q}\cdot J(\phi,\widetilde v)-{\Pr}^{-1}\cdot J(\psi,v) \end{equation}
\begin{equation} \label{eq59}
 \left(\frac{{\partial }}{{\partial t}}-\textrm{Pm}^{-1} \nabla ^2\right) \phi  - {\Pr}^{-1} \frac{{\partial \psi }}{{\partial z}}  =-{\Pr}^{-1} J(\psi ,\phi ))\end{equation}
\begin{equation} \label{eq60}
\left(\frac{{\partial \widetilde v}}{{\partial t}}-\textrm{Pm}^{-1} \nabla^2 \right)\widetilde v-{\Pr}^{-1} \frac{{\partial v}}{{\partial z}} +\textrm{Ro}\sqrt{\textrm{Ta}}\frac{{\partial \phi }}{{\partial z}}-2\xi{\Pr}^{-1}\textrm{Rb}\frac{\partial \psi}{\partial z}=$$
\begin{figure}
\centering
\includegraphics[width=5.5 cm, height=5.5 cm]{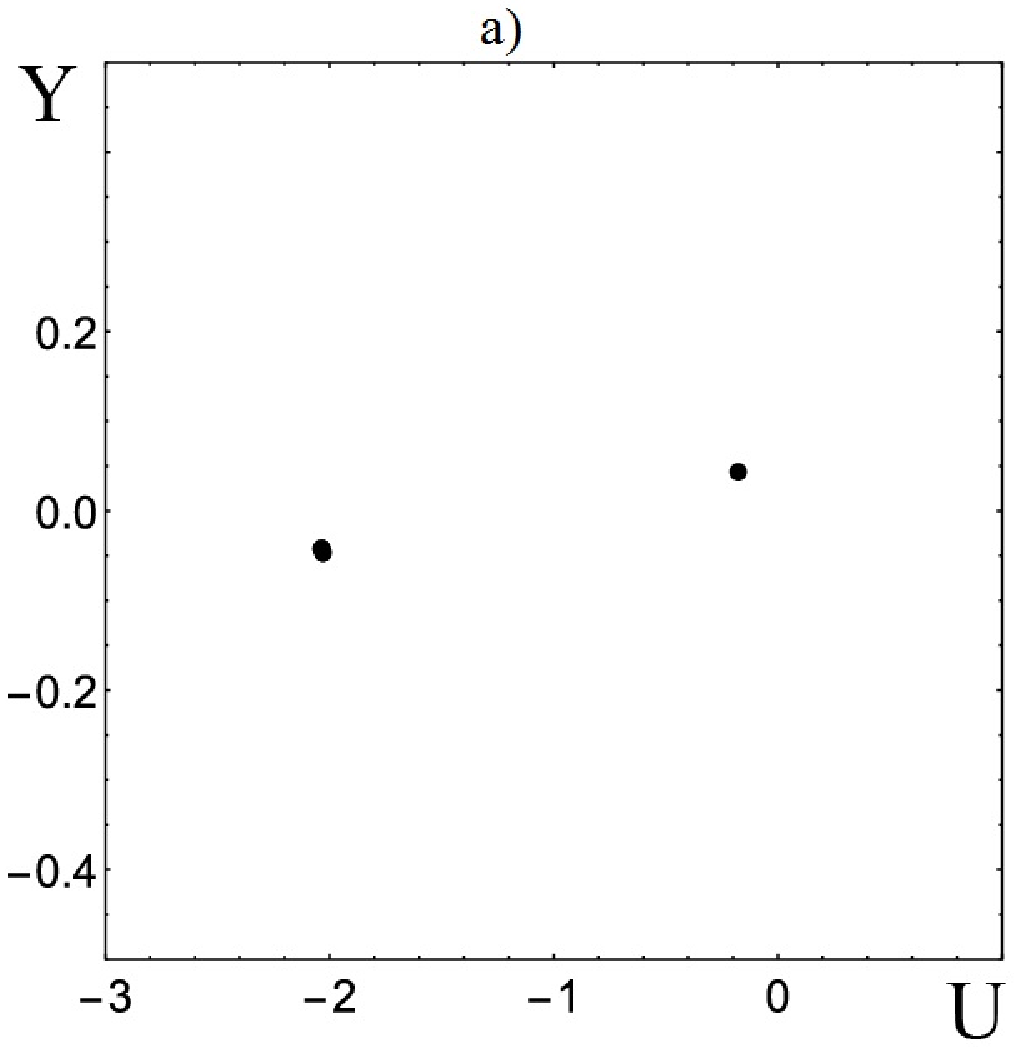} 
\includegraphics[width=5.5 cm, height=5.5 cm]{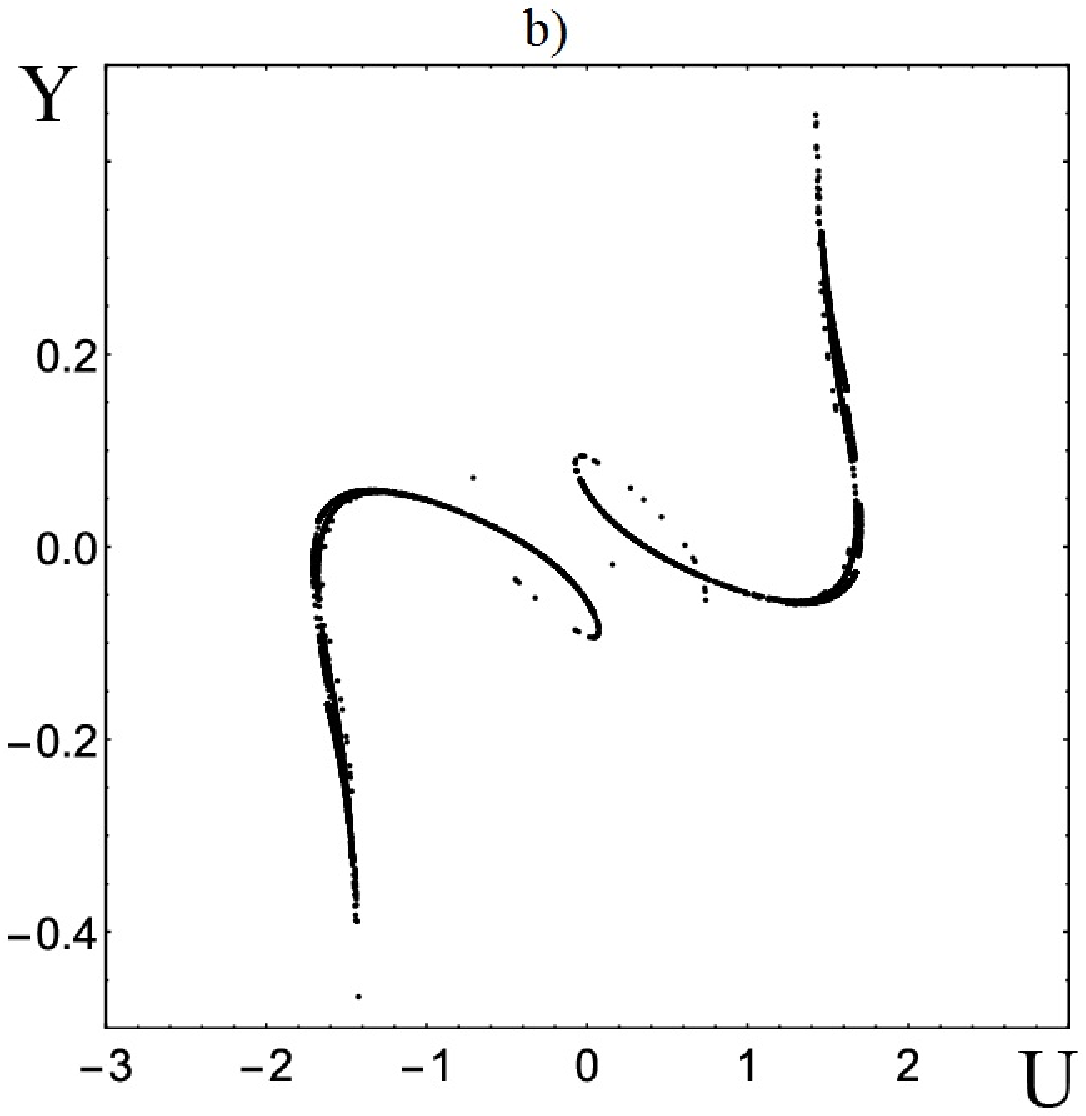}
\includegraphics[width=5.5 cm, height=5.5 cm]{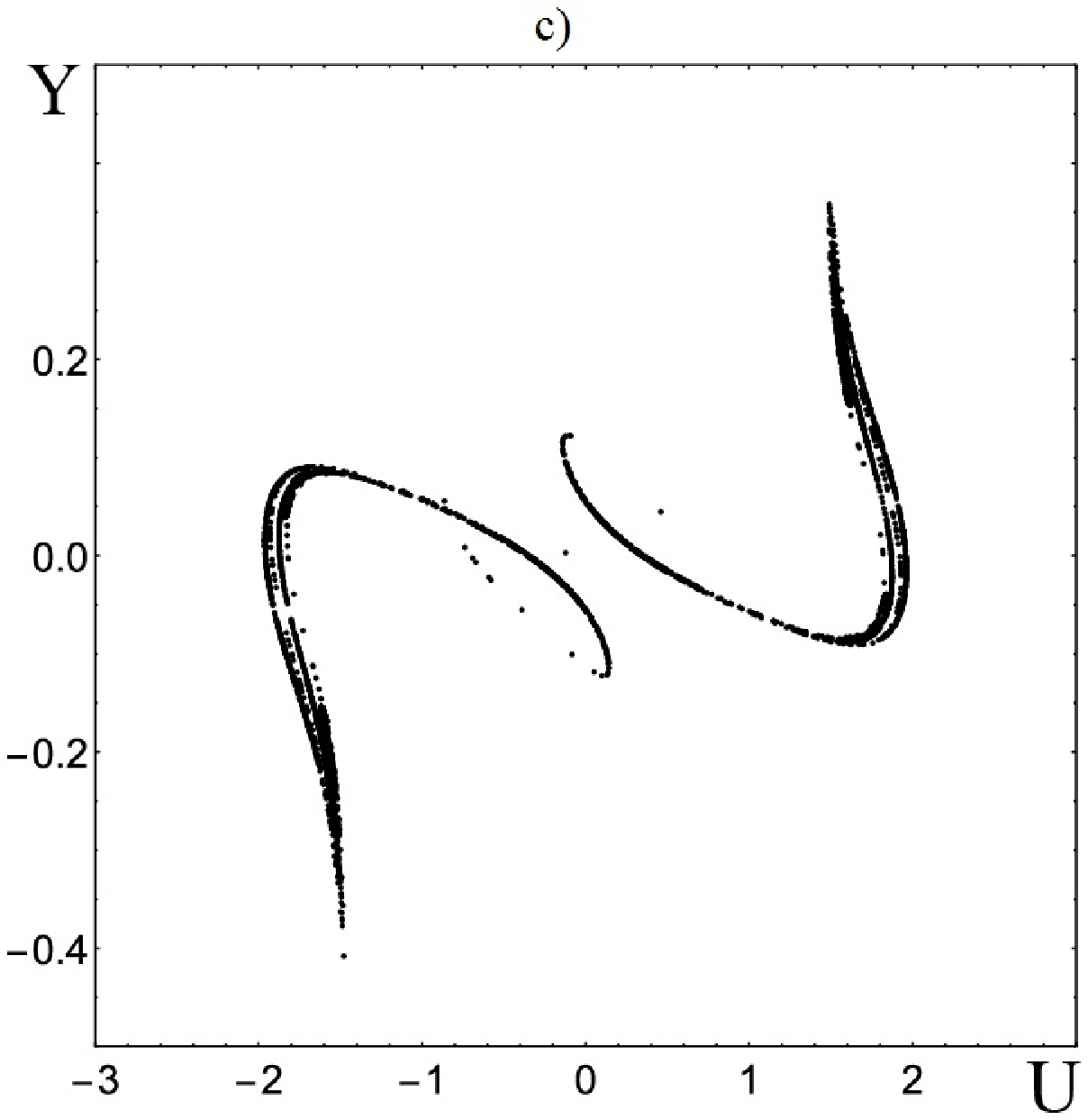} \\
\caption{ The Poincar\'{e} sections correspond to: a) the limit cycle at  $\textrm{R}=261.2$ (see Fig.  \ref{fg8}$\textrm{i}$-\ref{fg8}$\textrm{k}$); b) chaotic trajectories with $\textrm{R}=264.4$ (see Fig. \ref{fg8}$\textrm{l}$-\ref{fg8}$\textrm{m}$, Fig.  \ref{fg10}$\textrm{c}$ ); c) <<strong>> chaos at $\textrm{R}=330$ (see Fig. \ref{fg10}$\textrm{d}$ ).}\label{fg9}
\end{figure}
\begin{figure}
\centering
\includegraphics[width=19 cm, height=18 cm]{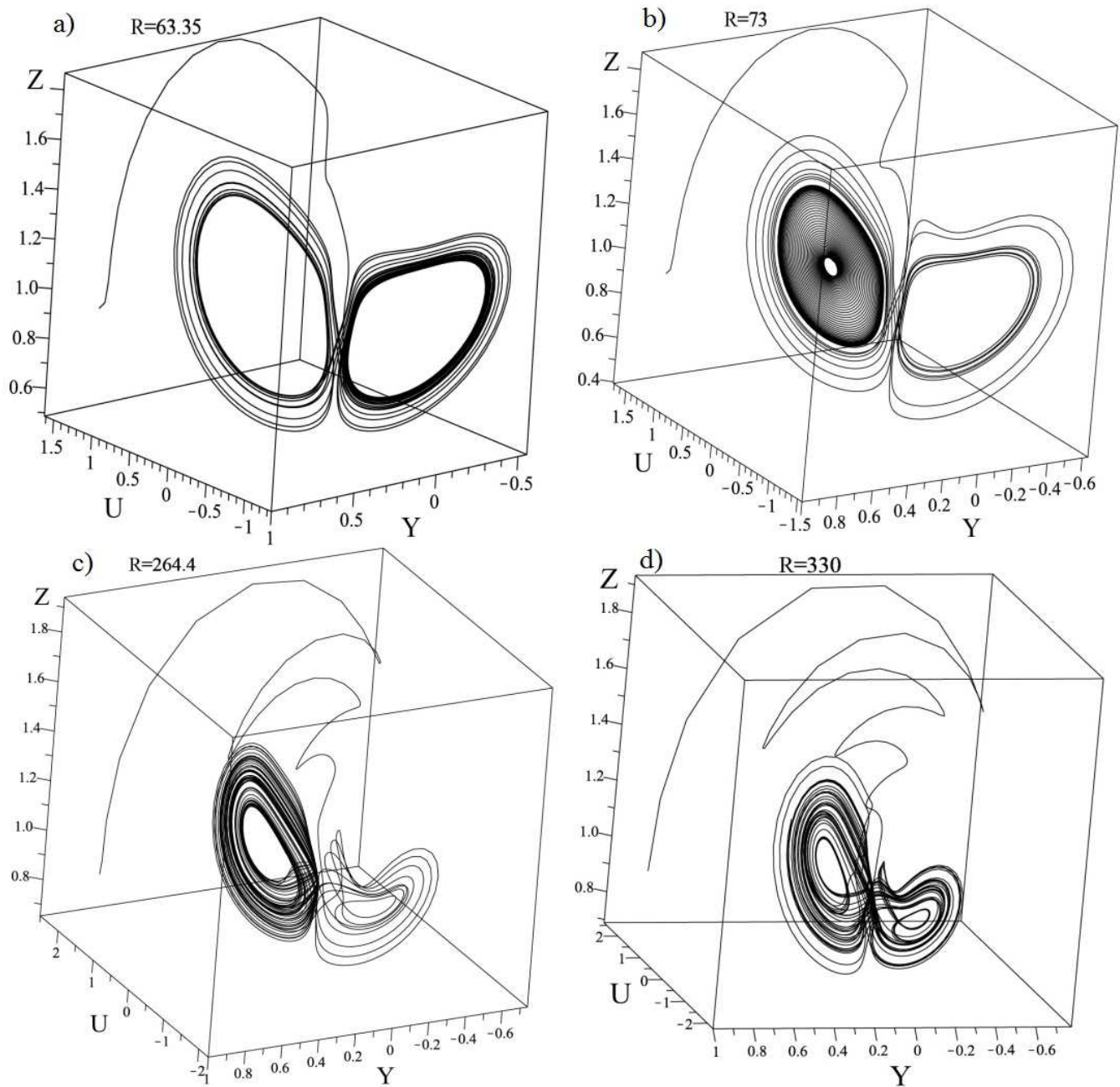} \\
\caption{The three-dimensional projections of the trajectories of chaotic motions with increasing parameter $\textrm{R}$. The upper figures $\textrm{a})$-$\textrm{b})$ correspond to the Rayleigh rotation profile $\textrm{Ro}=-1$, and the lower $\textrm{c})$-$\textrm{d})$ correspond to the Keplerian rotation profile $\textrm{Ro}=-3/4$. The magnetic Rossby number was considered equal to  $\textrm{Rb}=-1$.}\label{fg10}
\end{figure}
$$= {\Pr}^{-1}(J(\phi ,v) - J(\psi ,\widetilde v)) \end{equation}
\begin{equation} \label{eq61}
 \left(\frac{{\partial }}{{\partial t}}- \nabla^2 \right)\theta-{\Pr}^{-1}\frac{{\partial \psi }}{{\partial x}}=-{\Pr}^{-1} J(\psi,\theta))\end{equation}
The system of equations (\ref{eq57})-(\ref{eq61}) is complemented by the following boundary conditions:
\begin{equation} \label{eq62} 
\psi=\nabla^2\psi=\theta=\frac{d\phi}{dz}=0, \quad  \textrm{at} \quad z=(0,1) $$
$$\psi=\nabla^2\psi=v=\widetilde v=\phi=\frac{\partial \theta}{\partial x}= 0,\quad \textrm{at} \quad x=(0, L/h) \end{equation}
In the case when the external azimuthal magnetic field is absent $(\xi=0)$, the system of equations (\ref{eq57})-(\ref{eq61}) coincides with the results of \cite{39s}.

\subsection {Galerkin expansion}

To study the weakly nonlinear stage of the growth of convective instability, we use the Galerkin method, presenting all the perturbations in equations (\ref{eq57})-(\ref{eq61}) as a series taking into account boundary conditions (\ref{eq62}):
\[\label{eq63} \psi=\sum_{n=1}^{\infty}\sum_{m=1}^{\infty}A_{mn}(t)\sin(mkx)\sin(n\pi z) \]
\[v=\sum_{n=1}^{\infty}\sum_{m=1}^{\infty}V_{mn}(t)\sin(mkx)\cos(n\pi z) +\sum_{n=1}^{\infty}\sum_{m=1}^{\infty}\widetilde V_{mn}(t)\sin(mkx)\sin(n\pi z)\]
\begin{equation} 
\phi=\sum_{n=1}^{\infty}\sum_{m=1}^{\infty}B_{mn}(t)\sin(mkx)\cos(n\pi z) \end{equation}
\[\widetilde v=\sum_{n=1}^{\infty}\sum_{m=1}^{\infty}W_{mn}(t)\sin(mkx)\sin(n\pi z)+\sum_{n=1}^{\infty}\sum_{m=1}^{\infty}\widetilde W_{mn}(t)\sin(mkx)\cos(n\pi z) \]
\[\theta=\sum_{n=1}^{\infty}\sum_{m=1}^{\infty}C_{mn}(t)\cos(mkx)\sin(n\pi z) \]
\begin{figure}
  \centering
	\includegraphics[width=16 cm, height=8 cm]{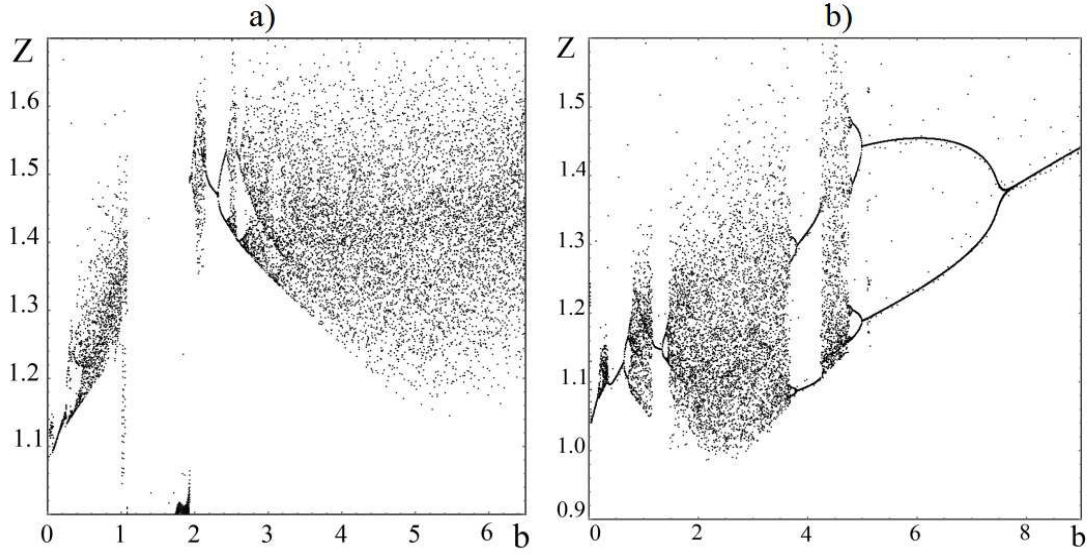} \\
	\caption{ Bifurcation diagrams for $Z$-amplitudes depend on changes in parameter $b$. The calculations were performed with the constant parameters  $\textrm{Pm}=1$, $\textrm{Pr}=9$, $\textrm{H}=5$, $\widetilde{\textrm{H}}=10$, $\widetilde{\xi}=1$, $\textrm{T}=1$, $\textrm{Ta}=2$, $\textrm{Rb}=-1$  for different rotation profiles $\textrm{Ro}$  and Rayleigh numbers $\textrm{R}$:  a) $\textrm{Ro}=-1$, $\textrm{R}=73$; b) $\textrm{Ro}=-3/4$, $\textrm{R}=330$.}\label{fg11}
\end{figure}
We limit our research with the Galerkin approximation of the minimal order, namely, for the current function $\psi$ we consider the modes  $(1,1)$, the $y$-components of the velocity $v$ -- $(1,1)+(1,1)$, $\phi$ of the magnetic potential -- $(1,1)$, $y$-components of the magnetic field perturbation $\widetilde v$ -- $(1,1)+(1,1)$ and temperature perturbations $\theta$ -- $(1,1)+(0,2)$:
\[ \psi(x,z,t)=A(t)\sin(kx)\sin(\pi z),\quad A(t)=A_{11}(t) \]
\[v=V(t)\sin(kx)\cos(\pi z)+\widetilde V(t)\sin(kx)\sin(\pi z), \quad V(t)=V_{11}(t), \quad \widetilde V(t)=\widetilde V_{11}(t)   \]
\begin{equation}\label{eq64} \phi(x,z,t)=B(t)\sin(kx)\cos(\pi z),\quad B(t)=B_{11}(t)  \end{equation}
\[\widetilde{v}=W(t)\sin(kx)\sin(\pi z) +\widetilde W(t)\sin(kx)\cos(\pi z),\quad W(t)=W_{11}(t), \quad \widetilde W(t)=\widetilde W_{11}(t) \]
\[\theta(x,y,t)=C_1(t)\cos(kx)\sin(\pi z)+C_2(t)\sin(2\pi z),\quad C_1(t)=C_{11}(t), \quad  C_2(t)=C_{02}(t), \]
where $k=2\pi h/L$  is the dimensionless wave number, $L$  is the characteristic length of the layer in the horizontal direction, $A$, $V$, $\widetilde V$, $B$, $W$, $\widetilde W$, $C_1$, $C_2$ are the disturbance amplitudes. As a result of substitution of the decomposition (\ref{eq64}) into equations (\ref{eq57})-(\ref{eq61}), and integration over the entire domain $[0,1]\times[0,L/h]$ , taking into account the orthogonality property
we get the evolution equations for perturbation amplitudes:
\begin{equation}\label{eq65} \frac{\partial A}{\partial\widetilde{t}}=-A-\frac{\pi\sqrt{\textrm{Ta}}}{a^4} V- \frac{\pi\textrm{QPr}}{a^2\textrm{Pm}} B+\frac{k\textrm{Ra}}{a^4} C_1+\frac{2\xi\pi\textrm{Q}\Pr}{a^4\textrm{Pm}} \widetilde W \end{equation}
\begin{equation}\label{eq66}
\frac{\partial V}{\partial\widetilde{t}}=-V+\frac{\pi\sqrt{\textrm{Ta}}}{a^2}(1+\textrm{Ro}) A+\frac{\pi\textrm{QPr}}{a^2\textrm{Pm}} W  \end{equation}
\begin{figure}
  \centering
	\includegraphics[width=16 cm, height=8 cm]{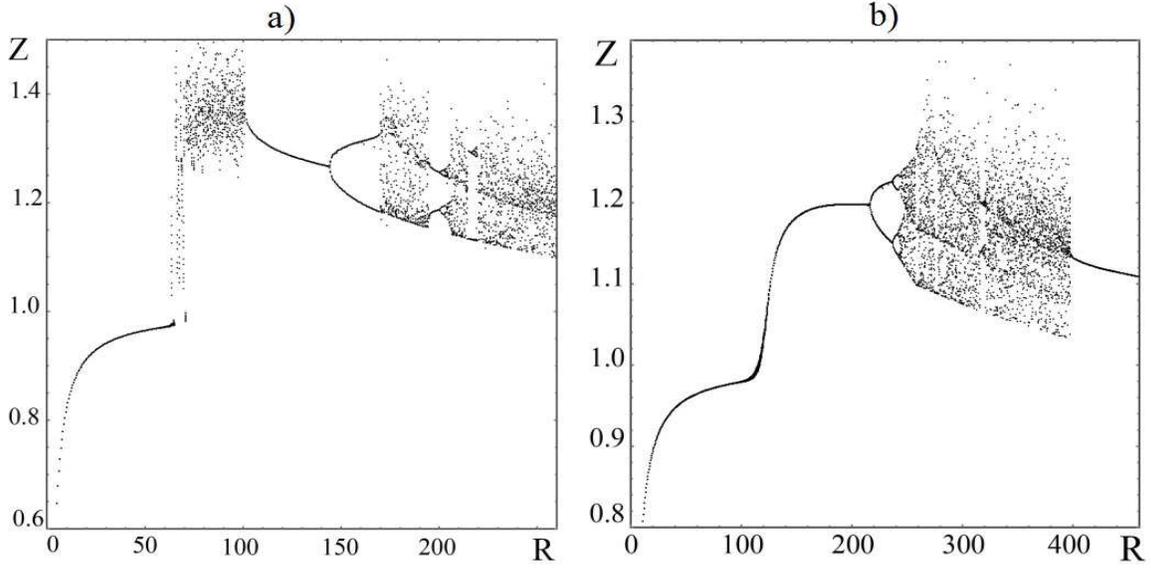} \\
	\caption{Bifurcation diagrams for $Z$-amplitudes depend on changes in the parameter $\textrm{R}$. The calculations were performed with the constant parameters $\textrm{Pm}=1, \textrm{Pr}=9, b=1, \textrm{H}=5, \widetilde{\textrm{H}}=10, \widetilde{\xi}=1, \textrm{T}=1, \textrm{Ta}=2$, $\textrm{Rb}=-1$ for different rotation profiles $\textrm{Ro}$: a) $\textrm{Ro}=-1$; b) $\textrm{Ro}=-3/4$.}\label{fg12}
\end{figure}
\begin{equation}\label{eq67}
\frac{\partial \widetilde V}{\partial\widetilde{t}}=-\widetilde V+\frac{2\xi\pi\textrm{Q}\Pr}{a^2\textrm{Pm}}(1+\textrm{Rb}) B-\frac{\pi\textrm{QPr}}{a^2\textrm{Pm}} \widetilde W  \end{equation}
\begin{equation}\label{eq68} \frac{\partial B}{\partial\widetilde{t}}=-\textrm{Pm}^{-1}B+\frac{\pi}{a^2}\textrm{Pr}^{-1} A  \end{equation}
\begin{equation}\label{eq69}  \frac{\partial W}{\partial\widetilde{t}}=-\textrm{Pm}^{-1}W-\frac{\pi}{a^2}\textrm{Pr}^{-1} V+\frac{\pi}{a^2}\textrm{Ro}\sqrt{\textrm{Ta}} B   \end{equation}
\begin{equation}\label{eq70}  \frac{\partial \widetilde W}{\partial\widetilde{t}}=-\textrm{Pm}^{-1}\widetilde W+\frac{\pi}{a^2}\textrm{Pr}^{-1}\widetilde V+\frac{2\xi\pi}{a^2}{\Pr}^{-1}\textrm{Rb} A  \end{equation}
\begin{equation}\label{eq71}  \frac{\partial C_1}{\partial\widetilde{t}}=-\textrm{Pr}^{-1} C_1+\frac{k}{a^2}\textrm{Pr}^{-1} A+\frac{\pi k}{a^2}\textrm{Pr}^{-1} A C_2  
\end{equation}
\begin{equation}\label{eq72}  \frac{\partial C_2}{\partial\widetilde{t}}=-\frac{4\pi^2}{a^2}\textrm{Pr}^{-1} C_2-\frac{\pi k}{2a^2}\textrm{Pr}^{-1} A C_1   \end{equation}
Here $a=\sqrt{k^2+\pi^2}$ is the total wave number and $\widetilde{t}=a^2t$ is the reduced time. The system of ordinary differential equations (\ref{eq65})-(\ref{eq72}) that we obtained is a low-order spectral model, but it can quite qualitatively reproduce convective processes in the nonlinear system of equations (\ref{eq57})-(\ref{eq61}). For convenience, we introduce the following notations
\[ \textrm{R}=\frac{k^2\textrm{Ra}}{a^6},\; \textrm{T}=\frac{\pi^2\sqrt{\textrm{Ta}}}{a^6}, \; \textrm{H}=\frac{\pi^2}{a^4}\frac{\textrm{QPr}}{\textrm{Pm}},\; b=\frac{4\pi^2}{a^2}, \; \widetilde{\textrm{H}}=\frac{2\xi \pi}{a^4}\textrm{H}, \; \widetilde{\xi}=\frac{2\xi a^2}{\pi}, \]
and rescale the amplitudes  $A$, $V$, $\widetilde V$, $B$, $W$, $\widetilde W$, $C_1$, $C_2$  in the form:
\[X(\widetilde{t})=\frac{k\pi}{a^2\sqrt{2}}A(\widetilde{t}),\; \mathcal{V}(\widetilde{t})=\frac{kV(\widetilde{t})}{\sqrt{2}},\;\widetilde{\mathcal{V}}(\widetilde{t})=\frac{k\widetilde V(\widetilde{t})}{\sqrt{2}},\; U(\widetilde{t})=\frac{kB(\widetilde{t})}{\sqrt{2}},\]
\begin{figure}
 \centering
\includegraphics[width=16 cm, height=12 cm]{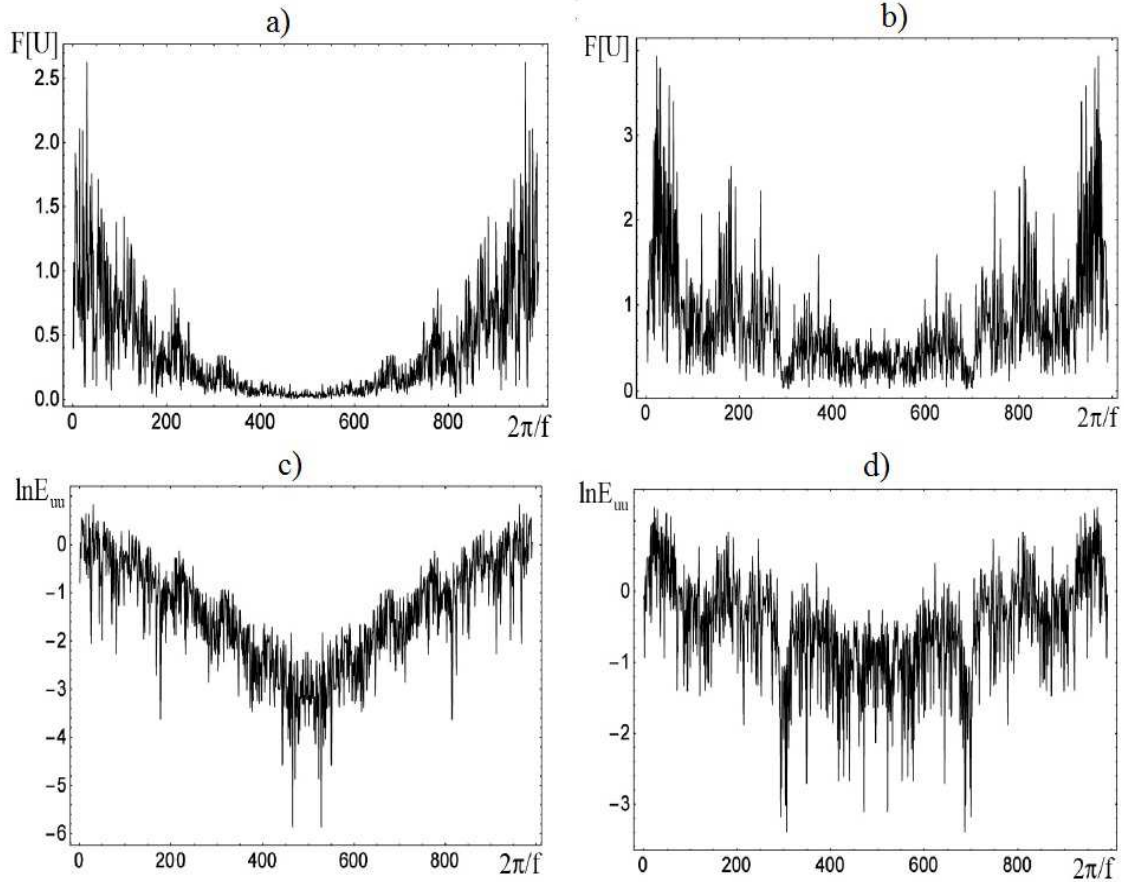} \\
\caption{Fourier spectra of magnetic field perturbations $U(\widetilde{t}) $: $\textrm{a)}$ $\textrm{Ro}=-1$ ($\textrm{R}=73$), $\textrm{b)}$ $\textrm{Ro}=-3/4$ ($\textrm{R}=330$); the Fourier spectra of the energy of a disturbed  magnetic field on a logarithmic scale: $\textrm{c)}$ $\textrm{Ro}=-1$ ($\textrm{R}=73$), $\textrm{d)}$ $\textrm{Ro}=-3/4$ ($\textrm{R}=330$).  The calculations were performed with the parameters $b=1$, $\textrm{Pm}=1, \textrm{Pr}=9, \textrm{H}=5, \widetilde{\textrm{H}}=10, \widetilde{\xi}=1, \textrm{T}=1,\textrm{Ta}=2, \textrm{Rb}=-1$.}\label{fg13}
\end{figure}
\[\mathcal{W}(\widetilde{t})=\frac{a^2k}{\pi\sqrt{2}}W(\widetilde{t}),\; \widetilde {\mathcal{W}}(\widetilde{t})=\frac{a^2k}{\pi\sqrt{2}}\widetilde W(\widetilde{t}),\; Y(\widetilde{t})=\frac{\pi C_1(\widetilde{t})}{\sqrt{2}}, \; Z(\widetilde{t})=-\pi C_2(\widetilde{t}) .\]
Then equations (\ref{eq65})-(\ref{eq72}) take the form of a nonlinear dynamic system of equations:
\begin{equation}\label{eq73}
\left\{
\begin{aligned}
\dot{X}=-X+\textrm{R}Y-\textrm{T}\mathcal{V}-\textrm{H}U+\widetilde{\textrm{H}}\widetilde{\mathcal{W}} \\
\dot{\mathcal V}=-\mathcal V+\textrm{H}\mathcal W+\sqrt{\textrm{Ta}}(1+\textrm{Ro})X    \\
\dot{\widetilde{\mathcal V}}=-\widetilde{ \mathcal V}+\widetilde{\xi} \textrm{H}(1+\textrm{Rb})U- \textrm{H}\widetilde{\mathcal W}   \\
\dot{U}=-\textrm{Pm}^{-1}U+\textrm{Pr}^{-1}X \\
\dot{\mathcal W}=-\textrm{Pm}^{-1}\mathcal W-\textrm{Pr}^{-1}\mathcal V+\textrm{Ro}\sqrt{\textrm{Ta}}U \\
\dot{\widetilde{\mathcal W}}=-\textrm{Pm}^{-1}\widetilde{\mathcal W}+\textrm{Pr}^{-1}\widetilde{\mathcal V}+\widetilde{\xi}\textrm{Pr}^{-1}\textrm{Rb}X \\
\dot{Y}=\textrm{Pr}^{-1}(-Y+X-XZ) \\
\dot{Z}=\textrm{Pr}^{-1}(-b Z+XY) \\
\end{aligned}
\right.
\end{equation}
where the dot above denotes time $\widetilde{t}$ differentiation. The last two nonlinear equations in system (\ref{eq73}) are similar to similar equations in the Lorentz system \cite{42s}-\cite{43s}. Therefore, we attribute the obtained nonlinear system of equations (\ref{eq73}) to equations of Lorentz type for eight-dimensional $(8D)$ phase space. In the limiting case, when there is only a constant external axial magnetic field, equations (\ref{eq73})  go over to the Lorentz equations for the six-dimensional $(6D)$  phase space, which were numerically investigated in \cite{39s}.

\subsection {Stability analysis}

In this section, we will conduct a qualitative \cite{44s} and numerical analysis of the dynamic system of equations (\ref{eq73}), which will allow us to determine the type of fixed points and the conditions for the occurrence of a chaotic regime. It is easy to note that the system of equations (\ref{eq73}) is homogeneous, since it lacks free terms. The trivial solution (\ref{eq73}), corresponding to the absence of convection, leads to the appearance of a special fixed point
\[O(X,\mathcal{V},\widetilde{\mathcal{V}},U,\mathcal{W},\widetilde{\mathcal{W}},Y,Z)=O(0,0,0,0,0,0,0,0), \]
which does not depend on the values of the parameters $\textrm{R}, \textrm{T}, \textrm{H},\textrm{Pm}, \textrm{Pr}, \widetilde{\xi}, b$. 
\begin{table}[h]
\centering
\begin{tabular}{|l|p{0.13\textwidth}|p{0.13\textwidth}|p{0.13\textwidth}|p{0.13\textwidth}|p{0.13\textwidth}|p{0.13\textwidth}|}
\hline $\textrm{R}$ & $\lambda_1$ & $\lambda_2$ & $\lambda_3$ & $\lambda_4$ & $\lambda_5$  & $\lambda_6$   \\ \hline
 $ 1.7$ &  $-1.018415$ $+i1.523391$  & $-0.003757$ &  $-0.111111$  & $-0.605587$ & $-1.464935$ & $-1.018415$  $-i1.523391$  \\   \hline
$ 5 $ &  $-0.111111$  & $0.170602 $   &  $-1.052476$ $+i1.464199$ & $ -0.668196$ & $-1.508564$ &   $-1.052476$ $-i1.464199$ \\   \hline
$17.3$ & $0.681274$  & $-1.147791$ $+i1.271764$ & $-0.111111$ & $-0.784116$ & $-1.712686$ & $-1.147791$ $-i1.271764$ \\ \hline
$ 17.4$ & $0.684900$ & $-1.148322$ $+i1.270355$ & $-0.1111111$ & $-0.784721$ & $-1.714645$ & $-1.148322$ $-i1.270355$ \\ \hline
$ 63.34277 $ & $1.930675$ & $-1.139159$ $+i0.910187$ & $ -0.111111$ & $-0.908022$ & $-2.855446$ & $-1.139159$ $-i0.910187$ \\ \hline
$63.35 $ & $1.930830$ & $-1.139147$  $+i0.910165$ &  $-0.111111$ & $-0.908030$ & $-2.855615$ & $-1.139147$ $-i0.910165$ \\ \hline
$73 $ & $-0.111111$ & $2.130020$ &  $-1.124679$ $+i0.885030$ & $-0.918213$ & $-3.073559$ & $-1.124679$ $-i0.885030$ \\ \hline
\end{tabular}
\caption{Eigenvalues $\lambda_{1,2,3,4,5,6}$ (Lyapunov indices) for the fixed point $O_1$, calculated for different values of the parameter $\textrm{R}$ under the condition: $\textrm{Pm}=1, \textrm{Pr}=9, \textrm{H}=5,\widetilde{\textrm{H}}=10, \widetilde{\xi}=1, \textrm{T}=1, \textrm{Ta}=2, b=1, \textrm{Rb}=-1 $ for the Rayleigh rotation profile $\textrm{Ro}=-1$.}
\label{tab1}
\end{table} 
The system of equations (\ref{eq73}) is invariant with respect to the replacement
$$(X,\mathcal{V},\widetilde{ \mathcal{V}},U,\mathcal{W},\widetilde{\mathcal{W}},Y,Z)\rightarrow (-X,-V,-\widetilde{ \mathcal{V}},-U,-\mathcal{W},-\widetilde{ \mathcal{W}},-Y,Z)$$
and dissipative, since the divergence of the vector field is negative:
\[\textrm{div}{\bf{\Phi}}=\frac{\partial \dot{X} }{\partial X}+\frac{\partial \dot{\mathcal{V}} }{\partial \mathcal{V}}+\frac{\partial \dot{\widetilde{ \mathcal{V}}} }{\partial \widetilde{ \mathcal{V}}}+\frac{\partial \dot{U} }{\partial U}+\frac{\partial \dot{\mathcal{W}} }{\partial \mathcal{W}}+\frac{\partial \dot{\widetilde{ \mathcal{W}}} }{\partial \widetilde{ \mathcal{W}}}+\frac{\partial \dot{Y} }{\partial Y}+\frac{{\partial \dot{Z}} }{\partial Z}=\]
\[-3(1+\textrm{Pm}^{-1})-\textrm{Pr}^{-1}(1+b)<0 \]
Due to dissipation, the phase volume shrinks:
\[{\bf{\Phi}}(\widetilde{t})={\bf{\Phi}}(0)\exp[\left(-3(1+\textrm{Pm}^{-1})-\textrm{Pr}^{-1}(1+b)\right)\widetilde{t}]. \]
Consequently, in the phase space of dissipative systems, attracting sets can arise - attractors. Equating the left sides of equations (\ref{eq73}) to zero, we get three equilibrium states:
\begin{equation}\label{eq74}
O_1(X_1,\mathcal{V}_1,\widetilde{\mathcal{V}}_1,U_1,\mathcal{W}_1,\widetilde{\mathcal{W}}_1,Y_1,Z_1), \quad O_2(X_2,\mathcal{V}_2,\widetilde{\mathcal{V}}_2,U_2,\mathcal{W}_2,\widetilde{\mathcal{W}}_2,Y_2,Z_2),$$
$$\quad O_3(X_3,\mathcal{V}_3,\widetilde{\mathcal{V}}_3,U_3,\mathcal{W}_3,\widetilde{\mathcal{W}}_3,Y_3,Z_3), \end{equation}
where the fixed points $X_{1,2,3},\mathcal{V}_{1,2,3},\widetilde{\mathcal{V}}_{1,2,3}, U_{1,2,3},\mathcal{W}_{1,2,3},\widetilde{\mathcal{W}}_{1,2,3},Y_{1,2,3},Z_{1,2,3}$  are respectively equal to:
\[(X_1,\mathcal{V}_1,\widetilde{\mathcal{V}}_1,U_1,\mathcal{W}_1,\widetilde{\mathcal{W}}_1,Y_1,Z_1)=(0,0,0,0,0,0,0,0),\]
\[\left(X_2,X_3\right)=\pm \frac{1}{r}\sqrt{b r ( \textrm{R}-r)},\; \left(\mathcal{V}_2,\mathcal{V}_3\right)=\pm \frac{\sqrt{\textrm{Ta}}\left(\textrm{HRo}\textrm{Pm}^2+\textrm{Pr}(1+\textrm{Ro})\right)}{r(\textrm{HPm}+\textrm{Pr})}\sqrt{b r (\textrm{R}-r)} ,  \]
\[\left(\widetilde{\mathcal{V}}_2,\widetilde{\mathcal{V}}_3\right)=\pm \frac{\widetilde{\xi}\textrm{HPm}}{r(\textrm{HPm}+\textrm{Pr})}\sqrt{b r (\textrm{R}-r)}, \; \left(U_2,U_3\right)=\pm \frac{\textrm{Pm}}{r\textrm{Pr}} \sqrt{b r (\textrm{R}-r)}, \]
\[ \left(\mathcal{W}_2,\mathcal{W}_3\right)=\pm \frac{\sqrt{\textrm{Ta}}\textrm{Pm}\left(\textrm{RoPm}-\textrm{Ro}-1\right)}{r(\textrm{HPm}+\textrm{Pr})} \sqrt{b r (\textrm{R}-r)},  \]
\[\left(\widetilde{\mathcal{W}}_2,\widetilde{\mathcal{W}}_3\right)=\pm \widetilde{\xi}\textrm{Pm}\cdot\frac{\textrm{HPm}+\textrm{Rb}(\Pr+\textrm{HPm})}{r\Pr(\textrm{HPm}+\textrm{Pr})} \sqrt{b r (\textrm{R}-r)},  \]
\[ \left(Y_2,Y_3\right)=\pm \frac{1}{\textrm{R}}\sqrt{b r (R-r)},\;\left(Z_2,Z_3\right)=1-\frac{r}{\textrm{R}}, \]
where
\[r= 1+\frac{\textrm{Pm}}{\textrm{Pr}}\textrm{H}+\frac{\textrm{T}\sqrt{\textrm{Ta}}\left[1+\textrm{Ro}\left(1+\frac{\textrm{Pm}^2}{\textrm{Pr}}\textrm{H}\right)\right]-\widetilde{\xi}\,\widetilde{\textrm{H}}\textrm{H}\frac{\textrm{Pm}^2}{\textrm{Pr}^2} }{1+\frac{\textrm{Pm}}{\textrm{Pr}}\textrm{H}}-\widetilde{\xi}\,\widetilde{\textrm{H}}\textrm{Rb}\frac{\textrm{Pm}}{\textrm{Pr}} \] 
\begin{figure}
 \centering
\includegraphics[width=19 cm, height=16 cm]{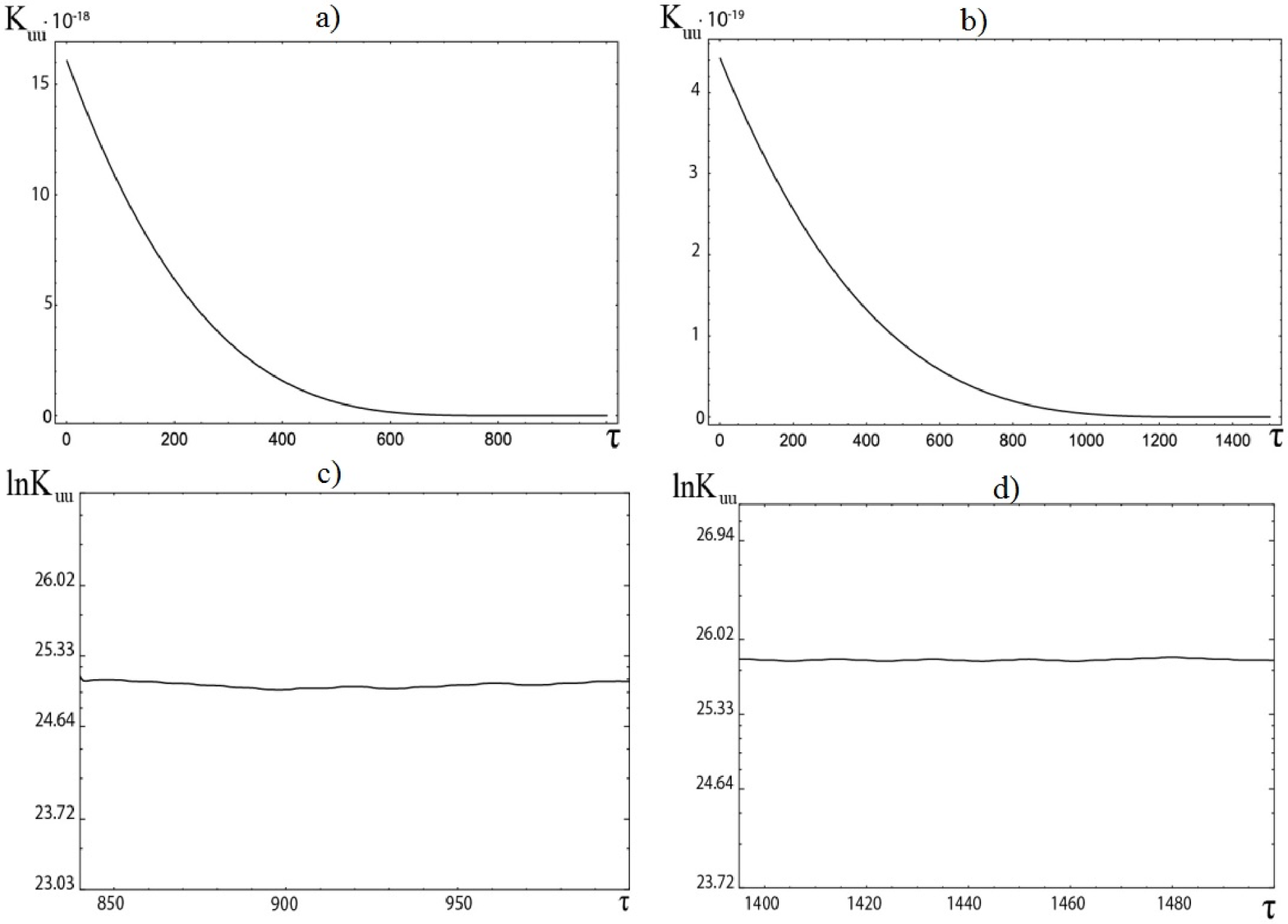}\\
\caption{Charts of the autocorrelation function $K_{UU}$  that depends on the time $\tau$ for chaotic magnetic field pulsations with parameters: a) $\textrm{R}=73$,  $b=1$, $\textrm{Pm}=1, \textrm{Pr}=9, \textrm{H}=5, \widetilde{\textrm{H}}=10, \widetilde{\xi}=1, \textrm{T}=1, \textrm{Ta}=2, \textrm{Rb}=-1,  \textrm{Ro}=-1$; b) $\textrm{R}=330$,  $b=1$, $\textrm{Pm}=1, \textrm{Pr}=9, \textrm{H}=5, \widetilde{\textrm{H}}=10, \widetilde{ \xi}=1, \textrm{T}=1, \textrm{Ta}=2, \textrm{Rb}=-1,  \textrm{Ro}=-3/4$; the graphs c)-d) correspond to the graphs for the autocorrelation functions $K_{UU}$ a)-b) in logarithmic scales from the time interval $\tau$. } \label{fg14}
\end{figure}
-- Rayleigh parameter for stationary convection: $r=\textrm{R}_{st}$,  which correspond to the critical Rayleigh numbers $\textrm{Ra}_{cr}$, which coincide with formula (\ref{eq42}). If the Rayleigh parameter $\textrm{R}=r$, then there is one fixed point $O_1$  in the system (no convection). With the Rayleigh parameters   $\textrm{R}>r$, the mode of the start of convection begins. The fixed points $O_2$  and  $O_3$ are symmetric to each other, and with the Rayleigh parameters $\textrm{R}<r$, their coordinates become imaginary.
We proceed to the study of stability, found equilibrium states (\ref{eq74}). To do this, linearize the system of equations (\ref{eq73}) in a small area of fixed points. Representing all variables in the form
\begin{table}[h]
\centering
\begin{tabular}{|l|p{0.13\textwidth}|p{0.13\textwidth}|p{0.13\textwidth}|p{0.13\textwidth}|p{0.13\textwidth}|p{0.13\textwidth}|}
\hline $\textrm{R}$ & $\lambda_1$ & $\lambda_2$ & $\lambda_3$ & $\lambda_4$ & $\lambda_5$  & $\lambda_6$   \\ \hline
 $ 7$ &  $-0.111111$  & $0.300721$   &  $-1.069207$ $+i1.373933$ & $-0.640047$ & $-1.633370$ & $-1.069207$  $ -i1.373933 $  \\   \hline
$ 25.4 $ &  $0.999853$  & $-1.161257$ $+i1.130827$  &  $-0.111111 $  & $ -0.785549$ & $ -2.002901 $ &   $ -1.161257 $ $-i1.130827$ \\   \hline
$25.5$ & $ 1.003024$  & $-1.161412$ $+i1.129763$ & $-0.111111$ & $-0.786012$ & $-2.005298$ & $-1.161412$ $-i1.129763$ \\ \hline
$157.3$ & $3.520377$ & $-1.060234$ $+i0.804160$ & $-0.111111$ & $-0.946442$ & $-4.564576$ & $-1.060234$ $-i0.804160$ \\ \hline
$ 261.2 $ & $ 4.743610 $ & $-1.037344$ $+i0.779631$ & $ -0.111111$ & $-0.966679$ & $ -5.813352$ & $-1.037344$ $-i0.779631 $ \\ \hline
$264.4 $ & $4.776903 $ & $ -1.036910 $ $+i0.779195$ & $-0.111111$ & $-0.967063$ & $ -5.847129 $ & $ -1.036910$  $-i0.779195 $ \\ \hline
$330 $ & $-0.111111$ & $ 5.419105 $ &  $-1.029801 $ $+i0.772189 $ & $-0.973367 $ & $-6.497246$ & $ -1.029801$ $-i0.772189 $ \\ \hline
\end{tabular}
\caption{ Eigenvalues $\lambda_{1,2,3,4,5,6}$ (Lyapunov indices) for the fixed point $O_1$, calculated for different values of the parameter $\textrm{R}$ under the condition: $\textrm{Pm}=1, \textrm{Pr}=9, \textrm{H}=5,\widetilde{\textrm{H}}=10, \widetilde{\xi}=1, \textrm{T}=1, \textrm{Ta}=2, b=1, \textrm{Rb}=-1 $ for the Keplerian rotation profile $\textrm{Ro}=-3/4$.}
\label{tab2}
\end{table} 
\[ \begin{pmatrix} X \\ \mathcal{V} \\ \widetilde{\mathcal{V}} \\ U\\ \mathcal{W} \\ \widetilde{ \mathcal{W}} \\ Y \\ Z \end{pmatrix}= \begin{pmatrix} X_0 \\ \mathcal{V}_0 \\ \widetilde{\mathcal{V}}_0 \\ U_0\\ \mathcal{W}_0 \\ \widetilde{\mathcal{W}}_0 \\ Y_0 \\ Z_0 \end{pmatrix}+ \begin{pmatrix} X^{'} \\ \mathcal{V}^{'} \\ \widetilde{\mathcal{V}}^{'} \\ U^{'}\\ \mathcal{W}^{'} \\ \widetilde{\mathcal{W}}^{'} \\ Y^{'} \\ Z^{'} \end{pmatrix} \cdot e^{\lambda \widetilde t}   \]
we obtain a linear algebraic system:
\begin{equation}\label{eq75}
\left\{
\begin{aligned}
(\lambda+1)X^{'}=\textrm{R}Y^{'}-\textrm{T}\mathcal{V^{'}}-\textrm{H}U^{'}+\widetilde{\textrm{H}}\widetilde{\mathcal{W^{'}}} \\
(\lambda+1)\mathcal V^{'}=\textrm{H}\mathcal W^{'}+\sqrt{\textrm{Ta}}(1+\textrm{Ro})X^{'}    \\
(\lambda+1)\widetilde{\mathcal V^{'}}=\widetilde{\xi} \textrm{H}(1+\textrm{Rb})U^{'}-\textrm{H}\widetilde{\mathcal W^{'}}   \\
(\lambda+\textrm{Pm}^{-1})U^{'}=\textrm{Pr}^{-1}X^{'} \\
(\lambda+\textrm{Pm}^{-1})\mathcal W^{'}=-\textrm{Pr}^{-1}\mathcal V^{'}+\textrm{Ro}\sqrt{\textrm{Ta}}U^{'} \\
(\lambda+\textrm{Pm}^{-1})\widetilde{\mathcal W^{'}}=\textrm{Pr}^{-1}\widetilde{ \mathcal V^{'}}+\widetilde{\xi}\textrm{Pr}^{-1}\textrm{Rb}X^{'} \\
(\lambda+\textrm{Pr}^{-1})Y^{'}=\textrm{Pr}^{-1}(X^{'}-X_0Z^{'}-X^{'}Z_0) \\
(\lambda+b\textrm{Pr}^{-1})Z^{'}=\textrm{Pr}^{-1}(X_0Y^{'}+X^{'}Y_0) \\
\end{aligned}
\right.
\end{equation}
\begin{table}[h]
\centering
\begin{tabular}{|l|p{0.13\textwidth}|p{0.13\textwidth}|p{0.13\textwidth}|p{0.13\textwidth}|p{0.13\textwidth}|p{0.13\textwidth}|}
\hline $\textrm{R}$ & $\lambda_1$ & $\lambda_2$ & $\lambda_3$ & $\lambda_4$ & $\lambda_5$  & $\lambda_6$   \\ \hline
$ 1.7$ &  $0.007037 $  & $-1.019144 $ $+i1.522268 $  &  $-0.118406 $  & $-0.606874 $ & $ -1.465690$ & $-1.019144 $  $-i1.522268 $  \\   \hline
$ 5 $ &  $ 0.184668$  & $ -1.054454$ $+i1.461110$  &  $ -0.116355$  & $-0.670607 $ & $-1.511018 $ &   $ -1.054454$ $-i1.461110 $ \\   \hline
$17.3$ & $ 0.708689$  & $-1.151507$ $+i1.263082 $ & $-0.115532 $ & $-0.787512$ & $ -1.724850 $ & $-1.151507 $ $-i1.263082$ \\ \hline
$ 17.4$ & $ 0.712405$ & $-1.152032 $ $+i1.261637 $ & $-0.115530 $ & $ -0.788119$ & $-1.726912 $ & $ -1.152032$ $-i1.261637  $ \\ \hline
$ 63.34277 $ & $ 1.984625$ & $-1.135871$ $+i0.903364 $ & $-0.115283$ & $-0.910469 $ & $-2.909351 $ & $-1.135871 $ $-i0.903364$ \\ \hline
$63.35$ & $ 1.984782 $ & $-1.135859 $ $+i0.903343$ & $-0.115283$ & $-0.910477 $ & $ -2.909524$ & $-1.135859 $  $-i0.903343$ \\ \hline
$73 $ & $2.187761 $ & $-1.121433$ $+i0.879271$  &  $ -0.115270$  & $-0.920460$ & $-3.131384$ & $ -1.121433$ $-i0.879271 $ \\ \hline
\end{tabular}
\caption{ Eigenvalues $\lambda_{1,2,3,4,5,6}$  (Lyapunov indices) for fixed points $O_{2,3}$, calculated for different values of the parameter \textrm{R} with the condition: $\textrm{Pm}=1, \textrm{Pr}=9, \textrm{H}=5,\widetilde{\textrm{H}}=10, \widetilde{\xi}=1, \textrm{T}=1, \textrm{Ta}=2, b=1, \textrm{Rb}=-1 $ for the Rayleigh rotation profile $\textrm{Ro}=-1$. }
\label{tab3}
\end{table}
Here the index "0" denotes fixed points, the stroke "${'}$"  denotes small perturbations, and $\lambda$ is the increase increment.
The eigenvalues of the system of equations (\ref{eq75}) $\lambda$  are found from the solution of the characteristic equation obtained by equating the determinant to zero, i.e.
\begin{equation}\label{eq76}
M_0\cdot\left((M_0-\widetilde{\xi}\,\widetilde{\textrm{H}}\textrm{Rb}\textrm{Pr}^{-1})\cdot M_2+\textrm{R}\textrm{Pr}^{-1}(\lambda+\textrm{Pm}^{-1})\cdot M_3\right)+\left(\textrm{T}\sqrt{\textrm{Ta}}\cdot M_1-\widetilde{\xi}\,\widetilde{\textrm{H}}\textrm{H}\textrm{Pr}^{-2} \right)\cdot M_2=0,
\end{equation}
where the following notation is entered
\[ M_0=(\lambda+1)(\lambda+\textrm{Pm}^{-1})+\textrm{H}\textrm{Pr}^{-1}, \; M_1=(1+\textrm{Ro})(\lambda+\textrm{Pm}^{-1})^2+\textrm{HRo}\textrm{Pr}^{-1}, \]
\[ M_2=(\lambda+\textrm{Pr}^{-1})(\lambda+b\textrm{Pr}^{-1}) +X_0^2\textrm{Pr}^{-2},\; M_3= X_0Y_0\textrm{Pr}^{-1}-(\lambda+b\textrm{Pr}^{-1})(1-Z_0).\]
\begin{table}[h]
\centering
\begin{tabular}{|l|p{0.13\textwidth}|p{0.13\textwidth}|p{0.13\textwidth}|p{0.13\textwidth}|p{0.13\textwidth}|p{0.13\textwidth}|}
\hline $\textrm{R}$ & $\lambda_1$ & $\lambda_2$ & $\lambda_3$ & $\lambda_4$ & $\lambda_5$  & $\lambda_6$   \\ \hline
 $ 7$ &  $-0.050863 $ $+i0.225281$ & $-1.019933 $ $+i1.589956$   &  $ -0.728712$ & $ -1.351915$ & $-1.019933$  $-i1.589956$ & $-0.050863 $  $ -i0.225281 $  \\   \hline
$25.4$ & $ 0.081593$ $+i0.280051$ & $-1.071357$ $+i1.473291$  & $-0.806221 $ & $ -1.436472 $ & $-1.071357$ $-i1.473291$ & $0.081593$ $-i0.280051$  \\ \hline
$ 25.5$ & $0.082260$ $+i0.280154$ & $ -1.071624$  $+i1.472687$  & $-0.806521$ & $-1.436973 $ & $ -1.071624$  $-i1.472687$ & $0.0822606$  $-i0.280154$  \\ \hline
$ 157.3 $ & $0.199088$   & $1.279224$   & $ -1.150952$ $+i0.964579$ & $ -0.939567$ & $ -2.459061$  & $-1.150952$  $-i0.964579 $  \\ \hline
$261.2$ & $0.173281$  & $ 2.003480$ & $ -1.099176$ $+i0.864594$ & $ -0.961561$ & $ -3.239070$ & $-1.099176 $ $-i0.864594$ \\ \hline
$264.4 $ & $ 0.172875$ & $ 2.022903 $ &  $-1.098049$ $+i0.862913$  & $ -0.961990$ & $ -3.259912$ & $ -1.098049$ $-i0.862913 $  \\ \hline
$330 $ & $ 0.166544$ & $ 2.395626 $ &  $-1.079254$ $+i0.836510$  & $ -0.969085$ & $ -3.656798$ & $ -1.079254$ $-i0.836510 $  \\ \hline
\end{tabular}
\caption{ Eigenvalues $\lambda_{1,2,3,4,5,6}$  (Lyapunov indices) for fixed points $O_{2,3}$, calculated for different values of the parameter \textrm{R} with the condition: $\textrm{Pm}=1, \textrm{Pr}=9, \textrm{H}=5,\widetilde{\textrm{H}}=10, \widetilde{\xi}=1, \textrm{T}=1, \textrm{Ta}=2, b=1, \textrm{Rb}=-1 $ for the  Keplerian  rotation profile $\textrm{Ro}=-3/4$.}
\label{tab4}
\end{table} 
Here $X_0=(X_1,X_2,X_3),\;Y_0=(Y_1,Y_2,Y_3),\;Z_0=(Z_1,Z_2,Z_3)$  are the coordinates of fixed points. In the limiting case, when the external azimuthal magnetic field is absent $\widetilde{\xi}=0$ from equation  (\ref{eq76}) we obtain the result of work \cite{42s}. If we substitute the values of the three equilibrium states (\ref{eq74}) into equation (\ref{eq76}), we obtain the characteristic equations for the eigenvalues $\lambda$ (Lyapunov indices) in each of these states. Moreover, for the points $O_2$ and $O_3$, the characteristic equations coincide. All characteristic equations can be written as a sixth-order polynomial:
\[P(\lambda)\equiv a_0\lambda^6+a_1\lambda^5+a_2\lambda^4+a_3\lambda^3+a_4\lambda^2+a_5\lambda+a_6=0, \]
where ${a}_0=1>0$. 

We do not give the explicit form of the real coefficients $a_1,a_2,a_3,a_4,a_5,a_6$ due to a very cumbersome form. However, from the theory of asymptotic stability \cite{45s} the Rauss-Hurwitz criterion is known, according to which for the polynomial $P(\lambda)$ to have all roots with negative real parts, it is necessary and sufficient to fulfill the conditions:\\
1)~ all coefficients of the polynomial $P(\lambda)$  are positive: $a_{n} >0$, $n=1\div 6$; \\
2)~ there are inequalities for the Hurwitz determinants: $\Delta _{n-1} >0$, $\Delta _{n-3} >0$ \dots . \\
Obviously, when the Rauss-Hurwitz criterion is satisfied, the fixed points are stable, and their equilibrium positions are classified as stable nodes.

Let us perform a numerical analysis of equation (\ref{eq76}) for the fixed point $O_1$ in the case of the Rayleigh rotation profile $\textrm{Ro}=-1$. Choosing the values of the parameters $\textrm{Pm}=1$, $\textrm{Pr}=9$, $\textrm{H}=5$, $\widetilde{ \textrm{H}}=10$, $\widetilde{\xi}=1$, \textrm{T}=1, \textrm{Ta}=2, $b=1$ and $\textrm{Rb}=-1$, we calculate the eigenvalues $\lambda_i$ depending on changes in the Rayleigh parameter $\textrm{R}$. These results are shown in Tab.  \ref{tab1}, and for the case of the Keplerian rotation profile $\textrm{Ro}=-3/4$ in Tab. \ref{tab2}. This shows that for negative  $\textrm{Re}\lambda<0$, the trajectories enter the point $O_1$, i.e. correspond to stable proper directions, and for positive $\textrm{Re}\lambda>0$, the trajectories leave the point $O_1$, and therefore correspond to unstable proper directions. The steady state of convection $(\lambda=0)$  corresponds to the critical value of the parameter
\begin{equation}\label{eq77}
\textrm{R}_{1cr}=1+\frac{\textrm{Pm}}{\textrm{Pr}}\textrm{H}+\frac{\textrm{T}\sqrt{\textrm{Ta}}\left[1+\textrm{Ro}\left(1+\frac{\textrm{Pm}^2}{\textrm{Pr}}\textrm{H}\right)\right]-\widetilde{\xi}\,\widetilde{\textrm{H}}\textrm{H}\frac{\textrm{Pm}^2}{\textrm{Pr}^2} }{1+\frac{\textrm{Pm}}{\textrm{Pr}}\textrm{H}}-\widetilde{\xi}\,\widetilde{\textrm{H}}\textrm{Rb}\frac{\textrm{Pm}}{\textrm{Pr}},  \end{equation} 
or critical value of the Rayleigh number
\[ \textrm{Ra}_{cr}=\frac{a^6}{k^2} + \frac{\pi^2 a^2\textrm{Q}}{k^2} + \frac{\pi^2a^4 \textrm{Ta}}{k^2 (a^4 + \pi^2\textrm{Q})} + \frac{\pi^2  \textrm{TaRo}(a^4+\pi^2\textrm{QPm})-4\pi^4\xi^2 \textrm{Q}^2}{k^2(a^4+\pi^2\textrm{Q})}-\frac{4\pi^2}{k^2}\cdot\xi^2\textrm{QRb}, \]
coinciding with formula (\ref{eq42}) and the expression for $r$. For the above numerical values of the parameters, the critical number is $\textrm{R}_{1cr}\approx 1.77$. In the case of the Kepler profile of rotation ($\textrm{Ro}=-3/4$), the critical Rayleigh number is slightly higher than $\textrm{R}_{1cr}\approx 2.12$. If the Rayleigh parameter $\textrm{R}=\textrm{R}_{1cr}$, then there is one fixed point $O_1(X_1,U_1,Y_1,Z_1)$ in the system. Performing similar reasoning, we calculate the eigenvalues $\lambda_i$ depending on changes in the Rayleigh parameter $\textrm{R}$ for the second (third) equilibrium state $O_{2,3}$. The results of these calculations for the Rayleigh rotation profile $\textrm{Ro}=-1$ are shown in Tab. \ref{tab3}, and for the case of the Keplerian rotation profile $\textrm{Ro}=-3/4$ in Tab. \ref{tab4}. Here, negative eigenvalues of  $\textrm{Re}\lambda<0$  correspond to stable eigen directions, and positive $\textrm{Re}\lambda>0$ correspond to unstable ones. The steady state of convection $(\lambda=0)$ corresponds to the critical value of the parameter $\textrm{R}_{2cr}$, which turns out to be equal to the first critical value: $\textrm{R}_{2cr}= \textrm{R}_{1cr}$.

\section { Discussion of numerical results}

In this section, we present the results of numerical studies of a nonlinear system of equations (\ref{eq73}) with initial conditions $X(0)=\mathcal{V}(0)=\widetilde{\mathcal{V}}(0)=U(0)= \mathcal{W}(0)=\widetilde{\mathcal{W}}(0)=Y(0)=Z(0)=1$ in the time domain  $ \widetilde{t}\in [0,15000] $ for the Rayleigh $(\textrm{Ro}=-1)$ and Keplerian $(\textrm{Ro}=-3/4)$ rotation profiles. Fig. \ref{fg7}$\textrm{a}$ shows the case for $\textrm{R}<\textrm{R}_{1cr}$, i.e. convection does not occur, and the initial perturbed magnetic field decays (Fig. \ref{fg7}$\textrm{b}$). As can be seen from Fig. \ref{fg7}$\textrm{a}$-\ref{fg7}$\textrm{b}$, the initial state of the system approaches the origin (point $O_1$), which is both a local and global attractor. When the numbers  $\textrm{R}>\textrm{R}_{1cr}$, a loss of stability occurs, and convective motions arise in the system. With the parameter $\textrm{R}=5$ (Fig. \ref{fg7}$\textrm{c}$-\ref{fg7}$\textrm{d}$) and  $\textrm{R}=7$ (Fig. \ref{fg8}$\textrm{a}$-\ref{fg8}$\textrm{b}$) in the phase plane $U-Y$ around the fixed point $O_2$  we observe the appearance of spiral trajectories that will wind as the parameter $\textrm{R}$ increases. It is noticeable at the value of $\textrm{R}=17.3$ (Fig. \ref{fg7}$\textrm{e}$) for the Rossby number $\textrm{Ro}=-1$ and at $\textrm{R}=25.4$ (Fig. \ref{fg8}$\textrm{c}$) for the Rossby number $\textrm{Ro}=-3/4$. In this case, the magnitude of the perturbed magnetic field oscillates damping in amplitude (Fig. \ref{fg7}$\textrm{f}$, Fig. \ref{fg8}$\textrm{d}$). In this case, the eigenvalues $\lambda_i$  are complex quantities with a negative real part, and therefore we classify the fixed point as a stable focus. A slight increase in the Rayleigh parameter from $\textrm{R}=17.3$ to $\textrm{R}=17.4$ (Fig. \ref{fg7}$\textrm{g}$) and similarly from $\textrm{R}=25.4$ to $\textrm{R}=25.5$  (Fig. \ref{fg8}$\textrm{e}$) leads to a change of sign (direction) of the oscillating perturbed magnetic field, which also attenuates (Fig. \ref{fg7}$\textrm{h}$, Fig. \ref{fg8}$\textrm{f}$). Here the phase trajectories wind in a spiral around a fixed point $O_3$ located in the negative sector of the plane $U-Y$, and we also classify it as a stable focus. In Fig. \ref{fg7}$\textrm{i}$-\ref{fg7}$\textrm{k}$ with the parameters $\textrm{R}=63.34277$  and $\textrm{Ro}=-1$, the occurrence of a homoclinic loop in phase space is shown. A similar picture is observed with the parameters $\textrm{R}=157.3$ and $\textrm{Ro}=-3/4$ (Fig. \ref{fg8}$\textrm{g}$-\ref{fg8}$\textrm{h}$). At $\textrm{R}=65.35$ $  (\textrm{Ro}=-1)$ , a transition from a homoclinic trajectory to a chaotic motion is observed (see Fig. \ref{fg7}$\textrm{l}$-\ref{fg7}$\textrm{m}$). The transition to chaotic trajectories for different values of the Rayleigh parameter $\textrm{R}$  can be seen in the bifurcation diagram shown in Fig. \ref{fg12}. For the case of a Keplerian rotation profile $(\textrm{Ro}=-3/4)$, the transition to the chaotic regime at $\textrm{R}=264.4$  (see Fig. \ref{fg8}$\textrm{l}$-\ref{fg8}$\textrm{m}$) occurs through the limit cycle at $\textrm{R}=261.2$ (see Fig. \ref{fg8}$\textrm{i}$-\ref{fg8}$\textrm{k}$), where the disturbed magnetic field makes periodic oscillations. Two points on the  Poincar\'{e}  section correspond to this process (Fig. \ref{fg9}$\textrm{a}$). The chaotic trajectories with the Rayleigh parameters $\textrm{R}=264.4$ and $\textrm{R}=330$ are shown in the Poincar\'{e} sections Fig. \ref{fg9}$\textrm{b}$-\ref{fg9}$\textrm{c}$. On the bifurcation diagram   Fig. \ref{fg12}$\textrm{b}$ we see a transition to chaos through a cascade of doubling of bifurcations. In Fig. \ref{fg7}$\textrm{m}$ and Fig. \ref{fg8}$\textrm{m}$ shows non-regular oscillations, with an aperiodic change in amplitude and direction (inversion) of a perturbed magnetic field. As can be seen from Fig. \ref{fg12}, a further increase in the parameter $\textrm{R}$ contributes to the development of chaotic convection behavior, in which chaotic fractal structures are formed -- strange attractors (see Fig. \ref{fg10}).

Fig. \ref{fg11} shows bifurcation diagrams from which the emergence of a chaotic regime through a series of period doubling bifurcations is seen. Results are shown for $Z$-amplitude versus $b$. A comparison of the bifurcation diagrams shows that the chaotic regime in the system occurs at lower values of the Rayleigh parameter ($\textrm{R}=45$) for the case of the Rayleigh rotation profile  $(\textrm{Ro}=-1)$. Both bifurcation diagrams show that with the growth of the parameter $b$, a large number of complex cycles occur, resulting in a chaotic (turbulent) state in the system.

Using spectral analysis of the system of equations (\ref{eq73}) by a numerical method, we obtained the complex aperiodic behavior of magnetic field perturbations with a noise-like frequency spectrum. Fig. \ref{fg13}$\textrm{a}$-\ref{fg13}$\textrm{b}$ shows the dependence of the Fourier components of the magnetic field perturbations $F[U]$ on the frequency $f$ for the Rayleigh numbers $\textrm{R}$ and Rossby $\textrm{Ro}$, respectively: $(\textrm{a})$ $\textrm{R}=73$, $\textrm{Ro}=-1$; $(\textrm{b})$ $\textrm{R}=330$, $\textrm{Ro}=-3/4$. Here you can see that the spectrum does not decrease with increasing frequency, and vice versa occurs after a certain zone of failure. Consequently, the solution obtained, with the above parameters, is really chaotic. Large bursts of the Fourier spectrum of the energy of a disturbed $\textrm{E}_{UU}$ magnetic field are observed with the Keplerian rotation profile $\textrm{Ro}=-3/4$ and the Rayleigh parameter $\textrm{R}=330$  than for the Rayleigh profile $\textrm{Ro}=-1$ and $\textrm{R}=73$ (see Fig. \ref{fg13}$\textrm{c}$-\ref{fg13}$\textrm{d}$).

The chaotic behavior of convection with different rotation profiles with Rossby numbers $\textrm{Ro}=-1$ and $\textrm{Ro}=-3/4$ is confirmed by the result of a numerical calculation of the autocorrelation function $K_{UU}(\tau)$ for magnetic field perturbations shown in Fig. \ref{fg14}. Here, the chaotic motion corresponds to the sections of the trajectories in Fig. \ref{fg14}$\textrm{a}$-\ref{fg14}$\textrm{b}$ with an exponential decay of the function $K_{UU}(\tau)$. Obviously, the exponential decay, on the logarithmic scale of the autocorrelation function $K_{UU}(\tau)$ is approximated by a straight line (see Fig. \ref{fg14}$\textrm{c}$-\ref{fg14}$\textrm{d}$).

\section {Conclusion}

In this work, linear and weakly non-linear Rayleigh-Benard convection of an inhomogeneously rotating electrically conducting fluid in a spiral magnetic field was investigated. In the linear approximation, we obtained the dispersion equation, from which, in the absence of a temperature gradient of $\textrm{Ra}=0$ and an external azimuthal magnetic field $\xi=0$ , the well-known criterion for the appearance of a standard MRI follows \cite{37s}. In the case when $\textrm{Ra}=0 $ and $\textrm{Ro}=0, \Omega=0 $  follows the stability criterion of an electrically conducting fluid in a toroidal (azimuthal) magnetic field \cite{37s}. The stationary and oscillatory modes of convection are considered depending on the profiles of inhomogeneous rotation (Rossby number $\textrm{Ro}$ ) and inhomogeneous azimuthal magnetic field (Rossby magnetic $\textrm{Rb}$ numbers). For stationary convection in a spiral magnetic field, the instability threshold increases towards positive Rossby numbers $\textrm{Ro}>0$. For oscillatory convective instability in a spiral magnetic field, the threshold Rayleigh number decreases with negative Rossby numbers $\textrm{Ro}<0$. A study of the chaotic mode of magnetic convection based on the equations of nonlinear dynamics of eight-dimensional phase space has been performed. We obtained these equations with the help of the Galerkin approximation of the minimal order. A qualitative analysis of a nonlinear system of dynamic equations is performed by analytical and numerical methods. As a result, the existence of a complex chaotic structure -- a strange attractor  was shown. In addition, a convection mode was found, in which a chaotic change of direction (inversion) and amplitude of the perturbed magnetic field occurs, taking into account the inhomogeneous rotation of the medium and the non-uniform external azimuthal magnetic field.

\end{document}